\documentclass[9pt,twocolumn,twoside]{pnas-new}

\usepackage{soul}
\usepackage{color}

 \newcommand{\rchem}{\bar{r}_{C}}
\templatetype{pnasresearcharticle} 
\setboolean{displaywatermark}{false}
\title{How kinesin waits for ATP affects the nucleotide and load dependence of the stepping kinetics}

\author[a]{Ryota Takaki}
\author[b]{Mauro L. Mugnai} 
\author[b]{Yonathan Goldtzvik }
\author[b]{D. Thirumalai }

\affil[a]{Department of Physics, The University of Texas at Austin, Austin, TX, USA}
\affil[b]{Department of Chemistry, The University of Texas at Austin, Austin, TX, USA}

\leadauthor{Takaki} 


\newcommand{\pN}{\mathrm{pN}}
\newcommand{\nm}{\mathrm{nm}}
\newcommand{\persec}{\mathrm{s}^{-1}}
\newcommand{\steps}{\mathrm{steps}}
\newcommand{\mMol}{\mathrm{mM}}
\newcommand{\uMol}{\mathrm{\mu M}}
\newcommand{\kp}{k^+}
\newcommand{\km}{k^-}

\newcommand{\Jp}{J^+}
\newcommand{\Jm}{J^-}
\newcommand{\Jgamma}{J^{\gamma}}

\begin{abstract}
Conventional Kinesin (Kin-1), which is responsible for directional transport of cellular vesicles, takes multiple nearly uniform 8.2 nm steps by consuming one ATP molecule per step as it walks towards the plus end of the microtubule (MT).
Despite decades of intensive experimental and theoretical studies there are gaps in the elucidation of key steps in the catalytic cycle of kinesin.
For example, how the motor waits for ATP to bind to the leading head has become controversial. 
Two experiments using a similar protocol, which follow the movement of a large gold nanoparticle attached to one of the motor heads, have arrived at different conclusions. 
One of them~\cite{Mickolajczyk_2015} asserts that kinesin waits for ATP in a state with both heads bound to the MT, whereas the other~\cite{isojima2016direct} shows that ATP binds to the leading head after the trailing head is detached. 
In order to discriminate between these two scenarios, we developed a minimal model, which analytically predicts the outcomes of a number of experimental observables quantities, such as the distribution of run length [$\boldsymbol{P(n)}$], the distribution of velocity [$\boldsymbol{P(v)}$], and the randomness parameter as a function of an external resistive force ($\boldsymbol{F}$) and ATP concentration ([T]).  
We find that $\boldsymbol{P(n)}$ is insensitive to the waiting state of kinesin. 
The bimodal velocity distribution $\boldsymbol{P(v)}$ depends on the ATP waiting states of kinesin. The differences in $\boldsymbol{P(v)}$ as a function of $\boldsymbol{F}$ between the two models may be amenable to experimental testing. 
Most importantly, we predict that the $\boldsymbol{F}$ and [T] dependence of the randomness parameters differ qualitatively depending on whether ATP waits with both heads bound to the MT or with detached tethered head. 
The randomness parameters as a function of $\boldsymbol{F}$ and [T]  can be quantitatively measured from stepping trajectories with very little prejudice in data analysis.  Therefore, an accurate measurement of the randomness parameter and the velocity distribution as a function of load and nucleotide concentration could resolve the apparent controversy, thus providing  insights into the waiting state of kinesin for ATP.  

\end{abstract}

\begin{document}

\maketitle
\thispagestyle{firststyle}
\ifthenelse{\boolean{shortarticle}}{\ifthenelse{\boolean{singlecolumn}}{\abscontentformatted}{\abscontent}}{}


\dropcap{K}inesin-1 (Kin1) is an archetypal cellular transporter, which moves along the microtubule (MT) to shuttle cargo towards the cellular periphery.
In the last quarter of century, a number of spectacular experimental studies~\cite{Svoboda93Nature,asbury2003kinesin,Block07BJ,mori2007kinesin,yildiz2004kinesin} have revealed many of the salient features of Kin1 structure and motility.
(i) Kin1 is a homodimer made up of two ATPase and MT-binding heads.
A key structural element, the neck-linker (NL) undergoes an order/disorder transition during the catalytic cycle termed ``NL docking". 
The distal tail forms a coiled coil which is responsible for dimerization and is also involved in cargo binding~\cite{MIKI2005467}.
(ii) Remarkably, the motor takes almost precisely 8.2 nm steps~\cite{yildiz2004kinesin}, which is commensurate with the spacing between two adjacent $\alpha \beta$ dimers -- the building blocks of the MT filament.
(iii) For each diffusional encounter with the MT, Kin1 takes multiple steps before detaching, a feature termed processivity~\cite{hackney1995highly}.
(iv) In the absence of resistive load ($F$), Kin1 moves nearly unidirectionally (backward steps are rare) towards the plus end of the MT~\cite{Pilling_2006}, and predominantly along a single protofilament~\cite{yildiz2008intramolecular}.
In addition, the velocity ($v$) distribution is roughly Gaussian with a peak typically in the range (100 - 1000) $\nm\cdot\persec$ depending on ATP concentration~\cite{Walter_2012}; 
the mean velocity is much larger than what is found in other motors such as Myosin V and Dynein.
As the resisting load increases, the probability that the motor takes backward steps becomes more prominent, reaching $0.5$ at the stall force $F_S \approx 7 \pN$~\cite{visscher1999single,carter2005mechanics}.
At stall, the mean motor velocity is zero, with a velocity distribution predicted to be bimodal and distinctly non-Gaussian~\cite{vu2016discrete}.
(v) The two heads step by a hand-over-hand mechanism~\cite{yildiz2004kinesin,asbury2003kinesin}, in which the trailing head (TH) detaches from the MT, bypasses the leading head (LH), and reattaches to the Target Binding Site (TBS) on the MT. Although it has long been advocated that 
the search for the TBS occurs largely by diffusion, it is only recently this has been definitively established~\cite{isojima2016direct,Zhang_2017,ZHANG2012628}.
The docking of the NL of the leading head (LH) propels the tethered head towards the + end of the MT, thereby minimizing the probability of taking backward steps. 
For this reason, NL docking is sometimes referred to as the ``power stroke''.
(vi) The energetic cost necessary to realize this directed motion is provided by the hydrolysis of ATP, which kinesin, like other motors, consumes parsimoniously. 
One molecule of ATP is hydrolyzed per step~\cite{schnitzer1997kinesin}.
The binding and hydrolysis of ATP are the events associated with the NL docking~\cite{asenjo2006nucleotide}. Based on these observations and other key experiments probing the variations of the stepping characteristics of the motor as a function of ATP concentration and applied load, several theoretical models for motors in general and the the catalytic cycle of Kin1 in particular have been proposed~\cite{fisher2001simple,liepelt2007kinesin,hwang2016quantifying,hwang2018energetic,sumi2017design,vu2016discrete,Wagoner16JPhysChemB}, although issues such as the mechanism of inter-head communication (gating) continue to be topics of interest~\cite{milic2014kinesin,andreasson2015examining,GENNERICH200959}. 

Despite these significant advances, there is a key problem related to the catalytic cycle of Kin1, which surprisingly still plagues the field: what is the waiting state of Kin1 for ATP binding? 
The answer to this fundamental question, which goes to one of the most important steps in the catalytic cycle of the motor, has been debated for nearly two decades, with contrasting pieces of evidence provided by optical trapping and single-molecule fluorescence experiments. Some studies have argued that the waiting state for ATP binding to the LH occurs when both the heads are bound (2HB) to MT~\cite{asenjo2003configuration}, whereas others assert that binding occurs only after the TH has detached, placing Kin1 in a one head bound (1HB) state~\cite{kawaguchi2001nucleotide,asenjo2009mobile}. The waiting state likely depends on ATP concentration. Kin1 waits with both heads bound (2HB) to MT at saturating ATP concentration whereas at low ATP concentration Kin1 might be in a one head bound state (1HB)~\cite{mori2007kinesin} before ATP binds.
However, in order to discriminate between the 1HB and 2HB ATP waiting states, it is necessary to monitor the location of the tethered head at the time of ATP binding, which requires experiments with high temporal and spatial resolution. 

The development of an experimental technique in which a large gold nanoparticle, AuNP, (between (20 - 40) nm in diameter) is attached to one of the heads, has made it possible to track indirectly the position of the tethered head during the stepping process as a function of ATP concentration. 
By tracking the location of the AuNP, either via interferometric scattering microscopy (iSCAT) ~\cite{Mickolajczyk_2015} or total internal-reflection dark-field microscopy~\cite{isojima2016direct}, two groups have achieved  a degree of temporal and spatial resolution necessary to resolve the waiting state of kinesin.  
From the analysis of the AuNP movement at different ATP concentrations, Micolajczky {\it et al.} argued that the motor waits in the 2HB state when the concentration of ATP is $\ge 10 \uMol$. 
The 2HB$\rightarrow$1HB transition follows ATP binding, and Kin1 spends about half of the stepping time with the tethered head parked above the bound head, which implies that the TH is displaced by about 8.2 nm from the initial binding site. 
In sharp contrast, Isojima {\it et al.} established that ATP binds to the LH only after the TH detaches from the MT. 
In other words, Kin1 waits for ATP in the 1HB state. 
In addition, computer simulations, using coarse-grained (CG) models for motors in general \cite{Alhadeff17PNAS,Mukherjee17PNAS} and kinesin in particular have provided insights into their functions. In particular, CG models  that accurately reproduce  several features found in experiments \cite{Hyeon11BJ,ZHANG2012628,Zhang_2017,Hyeon07PNAS,Hyeon07PNAS2}, have shown that the TH does spontaneously detach but does not walk towards the plus end of the MT until neck linker docks to the LH, which requires ATP binding to the leading head. 
These findings support the 1HB conformation as Kin1 ATP waiting state.
The contradictory findings reported in~\cite{Mickolajczyk_2015,isojima2016direct} and alluded to by Sindelar~\cite{sindelar2017tracking}, leave  the vexing question posed in the previous paragraph unanswered. This basic question needs to be fully answered in order to achieve a complete understanding of the stepping mechanism of conventional kinesin. 

It is unclear whether the differing conclusions reached in the recent experimental studies~\cite{Mickolajczyk_2015,isojima2016direct} arise because of the discrepancies in the constructs of the kinesin, the method of analysis of the trajectories, or due to the variations in the temporal resolution achieved in the experiments.  
Isojima {\it et al.} used a cys-lite motor in order to control the location of the linkage between the motor and the AuNP.
In contrast, Mickolajczyk {\it et al.} used a WT Kin1, whose N-terminus was extended with an Avi tag which is linked to the AuNP through a streptavidin-biotin complex.
Moreover, because of the higher temporal resolution in the dark field microscopy experiments~\cite{isojima2016direct}, Isojima {\it et al.} could discern the 1HB state by simultaneously monitoring the transverse fluctuations directly from the trajectories in a straightforward manner. 
On the other hand, Mickolajczyk {\it et al.}~\cite{Mickolajczyk_2015} relied on Hidden Markov Models (HMMs) to extract information from the stepping trajectories. 

In order to  discriminate between the contrasting interpretations of these experiments, it is desirable to  consider quantities that are straightforward to measure,
and that do not require indirect techniques of data analysis.
Ideally, a theoretical study capable of describing both scenarios (1HB and 2HB waiting state for ATP) should be able to identify which observable might be used to discriminate between the proposed cycles for kinesin.
Here, we use a simple and accurate model for kinesin stepping and calculate analytically a number of standard measurable quantities, such as the run length ($n$) distribution, $P(n)$, velocity ($v$) distribution, $P(v)$, and the randomness parameter as a function of ATP concentration (denoted as [T] from now on) and resistive force $F$. 
We show that $P(n)$ is independent of [T], and $P(v)$ as a function of [T] and $F$ are qualitatively similar for both the 1HB and 2HB models but differ quantitatively, a discrepancy that is amenable to experimental test. 
Remarkably, we predict that the mechanical and chemical randomness parameters, which are defined from readily measurable quantities, could be used to discriminate between the two scenarios. 
In particular, we find that both the mechanical and chemical randomness parameters at different [T] and $F$ are {\it qualitatively} different for these two scenarios in which Kin1 waits for ATP either in the 1HB or in the 2HB state.
Thus, we propose that measurements of the randomness parameters and $P(v)$ as a function of [T] and $F$ should unambiguously allow one to distinguish between the two very different ATP waiting states of Kin1.  

\section*{Results}
\begin{figure*}
\centering
\includegraphics[width=12.8 cm]{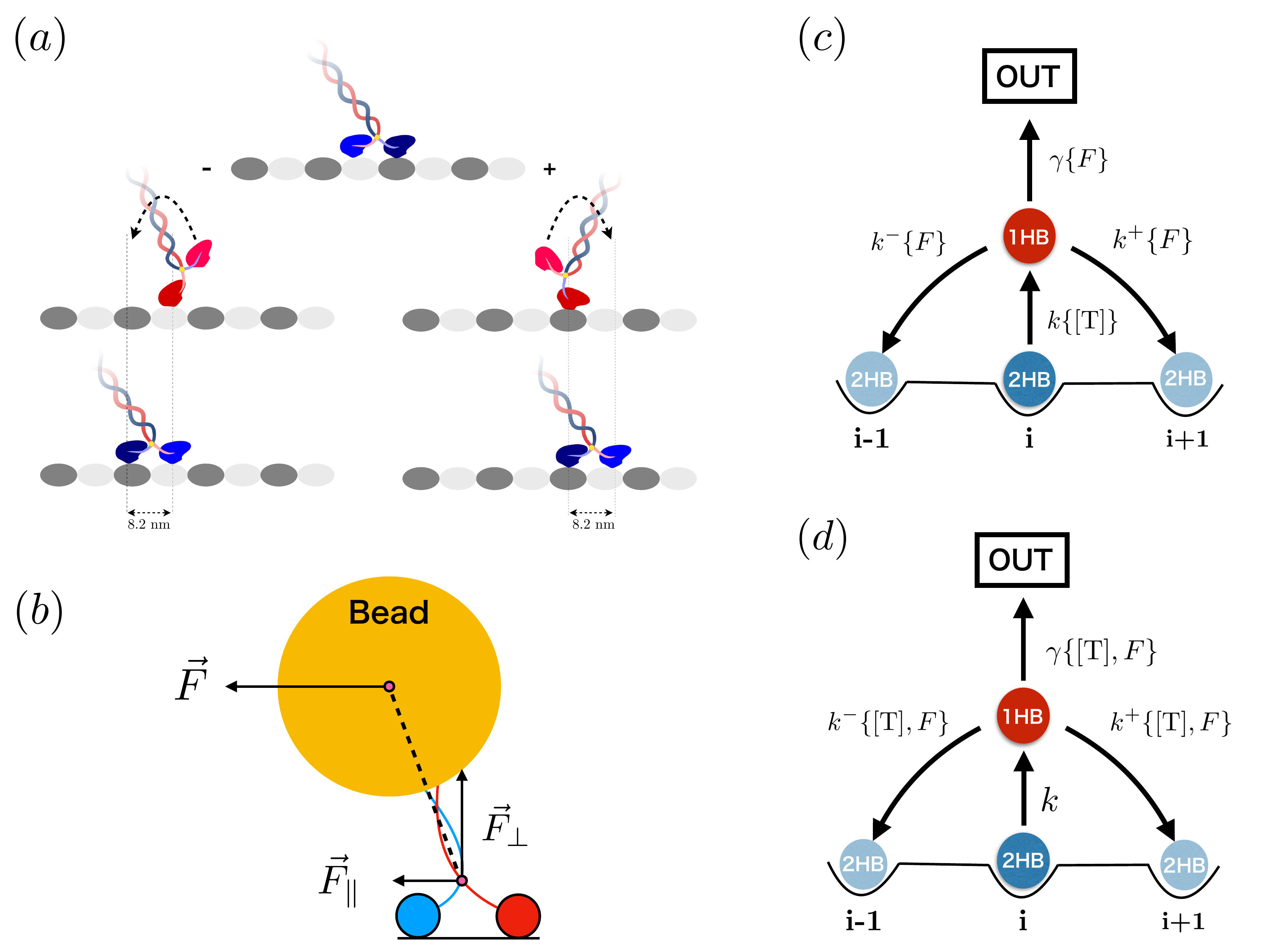}
\caption{\label{fig1} 
(a) Schematic representation of a kinesin motor walking hand over hand on the microtubule (MT). The tethered head detaches, undergoes diffusion, and passes the leading head (LH), and reattached to the target binding site that is roughly 16.4 nm from the starting position, resulting in a net displacement of 8.2 nm step. In the process  one ATP molecule is hydrolyzed.   
(b) Decomposition of the resistive force applied to the bead attached to the coiled coil of kinesin into perpendicular ($\perp$) and parallel ($\parallel$) direction to the MT. 
(c) Kinetic scheme  describing the ATP-waiting state showing that binding to the LH occurs when both the heads are bound to the MT in the 2HB state.
(d) Same as (c) except ATP binds when kinesin is in 1HB state. 
In both the scenarios the i-1 and i+1 state are equivalent to i in that they correspond to both the heads bound to the track. The difference is in the state that waits for ATP. The state labeled OUT represents an absorbing state. 
}
\end{figure*}
We begin by presenting some nomenclature.
We refer to the scenario in which ATP binds to the LH of kinesin  when both heads are attached to the MT as the ``2HB model'', whereas the ``1HB model'' refers to the alternative sequence of events, in which the detachment of the TH of kinesin precedes the binding of ATP to the LH.
In order to calculate $P(v)$ and $P(n)$ we created two versions of what is perhaps the simplest chemical kinetics model for a molecular motor [Fig.~\ref{fig1}(c) and (d)], one for the 2HB and one for the 1HB  model. 
The difference between the two lies in the the dependence on ATP concentration of the kinetic rates. 
In the 2HB model the transition to the 1HB state occurs only after ATP binds to the leading head [Fig.~\ref{fig1}(c)], therefore, the 2HB$\rightarrow$1HB rate accounts for the dependence on [T]. 
Because in the 1HB model ATP binds only after the tethered head detaches, the stepping rates, $k^+$ and $k^-$, as well as $\gamma$ are assumed to depend on [T] [Fig.~\ref{fig1}(d)].  

We use Michaelis-Menten kinetics to describe ATP binding and account for the effect of external load on the rates by adopting the Bell model.
In order to distinguish between the parallel component of the vectorial load applied to the motor, which introduce the symbols $\parallel$ and $\perp$, respectively [see Fig.~\ref{fig1}(b)].
For the 2HB model,  $k\{[\text{T}]\}=\frac{k_0[\text{T}]}{K_T+[\text{T}]}$,  $k^+(F)=k_0^+\text{e}^{-\beta F d^+}$, $k_0^-(F)=k^-\text{e}^{\beta F d^-}$, and $\gamma(F)=\gamma_0\text{e}^{|F|/F_d}$, where $d^\pm=d_\parallel^\pm F_\parallel/F$ and the load $F_d=(|F|k_BT)/(F_\perp d_\gamma)$. 
In the case of the 1HB model [Fig.~\ref{fig1}(d)], $k$ is a constant, independent of [T] and load, $k^+= \frac{k_0^+[\text{T}]}{K_T+[\text{T}]}\text{e}^{-\beta F d^+}$, $k^-= \frac{k_0^-[\text{T}]}{K_T+[\text{T}]}\text{e}^{\beta F d^-}$, and $\gamma= \frac{\gamma_0[\text{T}]}{K_T+[\text{T}]}\text{e}^{F/F_d}$. 
Note that in both the scenarios we have assumed that the 2HB$\rightarrow$1HB transition is independent of load. 

For the first step in the calculation of $P(v)$ and $P(n)$ we obtain the stationary fluxes for forward stepping, backward stepping, and detachment.
The motor is viewed as a random walker starting in the 2HB state at the MT site $i$. A steady-state probability distribution of occupying the 2HB and 1HB states is enforced by replenishing the 2HB state of all the walkers that step forward or backward (reaching $i+1$ and $i-1$, respectively) or detach~\cite{hillfree,Hill2879},
\begin{equation}
\begin{split}
\label{eq:ME}
\frac{dP_\text{2HB}}{dt}&=-kP_\text{2HB}+(\gamma+k^++k^-)P_\text{1HB}=0 \\
\frac{dP_\text{1HB}}{dt}&=-(\gamma+k^++k^-)P_\text{1HB} + kP_\text{2HB}=0,
\end{split}
\end{equation}    
The normalization condition implies that $P_\text{2HB} + P_\text{1HB} = 1$. 
The solution of Eq.(\ref{eq:ME}) gives $P_\text{1HB}=\frac{k}{k+k^++k^-+\gamma}$. 
The stationary fluxes for forward stepping ($\Jp$), backward stepping ($\Jm$), and detachment ($\Jgamma$) are computed by multiplying the steady-state probability of being in state 1HB ($P_\text{1HB}$) times $\kp$, $\km$, and $\gamma$~\cite{hillfree,Hill2879}, 
\begin{align} 
\begin{split}
\label{flux}
J^\pm &=\frac{k}{k_T}k^\pm,  \hspace{20pt}  J^\gamma =\frac{k}{k_T}\gamma,
\end{split}                 
\end{align}
where $k_T=k+k^++k^-+\gamma$.


The average velocity and run length are given by $V=s(J^+-J^-)$ and $L=V/J^{\gamma}$, respectively, where $s=8.2$ nm is the kinesin step size, which we assume is a constant. 
It is straightforward  to show that
\begin{align} 
\begin{split}
\label{VandL}
V=\frac{V_{\mathrm{max}}[\text{T}]}{K_{\mathrm{M}}+[\text{T}]},  \hspace{20pt}  L=\frac{(k^+-k^-)s}{\gamma},
\end{split}                 
\end{align}
for both the 2HB and 1HB model. 
The maximum velocity at saturating ATP concentration for the 2HB and 1HB model are given by  $V_{\mathrm{max}}=\frac{k_0(k^+-k^-)s}{k_0+k^++k^-+\gamma}$ and  $V_{\mathrm{max}}=(k_0^+\text{e}^{-\beta F_d^+} -k_0^-\text{e}^{\beta F_d^-})ks/k_T$, respectively. 
The concentrations at which the velocity of Kin1 is half-maximal are given by $K_{\mathrm{M}}=\frac{K_T(k^++k^-+\gamma)}{k_0+k^++k^-+\gamma}$ for the 2HB model [Fig.\ref{fig1}(c)] and $K_{\mathrm{M}}=K_T$ for 1HB model [Fig.\ref{fig1}(d)].

\subsection*{Run length distribution, $\boldsymbol{P(n)}$}
\ In order to solve for the run length and velocity distributions, we construct the joint probability [$P(m,l)$] that the motor takes $m$ forward steps and $l$ backward steps before detachment (see the Supplementary Information (SI) for details),
\begin{align} 
\begin{split}
\label{eq:Pml}
P(m,l)=\frac{(m+l)!}{m!l!}\Big(\frac{J^+}{J_T}\Big)^m\Big(\frac{J^-}{J_T}\Big)^l\Big(\frac{J^\gamma}{J_T}\Big).
\end{split}                 
\end{align}
In the above equation, $J^+/J_T$ ($J^-/J_T$) is the probability of taking a forward (backward) step starting from the 2HB state, and $J_T = J^+ + J^- + J^{\gamma}$.
Similarly, $J^{\gamma}/J_T$ is the probability that a motor in the 2HB state detaches.
The number of all the possible ways in which a sequence of $m$ forward and $l$ backward steps can be realized is accounted for by the binomial factor.
If the run length is $n=m-l$, then $P(n)$ is given by $P(n)=\sum_{m,l=0}^{\infty}P(m,l)\delta_{m-l,n}$,
where $\delta_{m-l,n}$ is the Kronecker delta function.
By carrying  out the summation we obtain,
\begin{align} 
\begin{split}
\label{eq:run_dist}
P(n\gtrless0)=\Big(\frac{2J^\pm}{J_T+\sqrt{J_T^2-4J^+J^-}}\Big)^{|n|}\frac{J^\gamma}{\sqrt{J_T^2-4J^+J^-}}.
\end{split}                 
\end{align}
Note that the functional form of $P(n\gtrless 0)$ is independent of the model considered -- it is the dependence of the fluxes on [T] and $F$ that separates the 2HB and 1HB model.
We note that the expression for $P(n)$ obtained here is equivalent to the one obtained previously\cite{vu2016discrete,zhang2017theoretical}, which can be  derived by substituting the rate of forward step, backward step, and detachment to the corresponding fluxes defined in Eq.(\ref{flux}).

\subsection*{Velocity distribution, $\boldsymbol{P(v)}$}
\ In order to calculate $P(v)$, we first compute $f(m,l,t)$, which is the joint probability density for detaching during a time interval from $t$ to $t+dt$ after the motor takes $m$ forward  and $l$ backward steps.
Let $f_+(t)$ be the probability density of taking a forward step between $t$ and $t+dt$, given that at $t=0$ the motor is in the 2HB state.
Similarly, the probability density for stepping backward and for detachment are denoted by $f_-(t)$ and $f_\gamma(t)$.
We show in the SI that $f_+(t)$, $f_-(t)$, and $f_\gamma(t)$ are linear combinations of two exponential functions with rates $\xi_1 = k$ and $\xi_2 = k^++k^-+\gamma$ (see Fig.~\ref{fig1} for the definition of the rates).
The probability density $f(m,l,t)$ is given by,
\begin{align}
\begin{split}
\label{eq:integralEq}
&f(m,l,t) = \frac{(m+l)!}{m!l!} \\
           &\int_0^t dt_{m+l} \int_0^{t_{m+l}} dt_{m+l-1} \dots \int_0^{t_3} dt_2\int_0^{t_2} dt_1 \\
           &\prod_{i=1}^{m} f_+(t_i-t_{i-1}) \prod_{i=m+1}^{m+l}f_-(t_i-t_{i-1})f_\gamma(t-t_{m+l}).
\end{split}
\end{align}
As detailed in the SI,  the solution of the integral equation in Eq.(\ref{eq:integralEq}) is,
\begin{equation}
\begin{split}
f(m,l,t) &= \frac{\gamma \sqrt{\pi}}{m!l!}e^{-\frac{\xi_1 + \xi_2}{2}t} t^{m+l} \frac{k^{m+l+1}(k^+)^{m}(k^-)^{l}}{|\xi_2 - \xi_1|^{m+l+1}} \\
           & \sqrt{|\xi_1-\xi_2|t} I_{m+l+\frac{1}{2}}\Big(\frac{|\xi_1-\xi_2|}{2}t\Big),
\end{split}
\label{eq:fmlt}
\end{equation}
where $I_{m+l+\frac{1}{2}}\Big(\frac{|\xi_1-\xi_2|}{2}t\Big)$ is the modified Bessel function of the first kind.
The velocity distribution may be obtained by changing the variables to $v = (m-l)/t$, which gives,
\begin{equation}
\begin{split}
P(v>0) =
& \sum_{\substack{m,l\\m>l}}^{\infty}\frac{m-l}{v^2} \frac{\gamma\sqrt{\pi}}{m!l!}\text{e}^{-\frac{\xi_1+\xi_2}{2}\frac{m-l}{v}}\big(\frac{m-l}{v}\big)^{m+l+\frac{1}{2}} \\
&  \frac{k^{m+l+1}(k^+)^m(k^-)^l}{|\xi_2-\xi_1|^{m+l+\frac{1}{2}}} I_{m+l+\frac{1}{2}}\big(\frac{|\xi_2-\xi_1|}{2}\frac{m-l}{v}\big).
\end{split}
\label{eq:pv}
\end{equation}
The expression for $P(v<0)$ is presented in the SI.
Note that both Eq.~(\ref{eq:fmlt}) and Eq.~(\ref{eq:pv}) hold if $\xi_1 \ne \xi_2$. 
However, as we show in the SI, that the solution for $\xi_1 = \xi_2$ has the same form, and can be obtained  as the limit for $\xi_1 \rightarrow \xi_2$ of Eq.~(\ref{eq:pv}).
Again, the functional form for $P(v)$ is the same in the 2HB and 1HB model, which are only differentiated by the dependence on $F$ and [T] of the chemical rates. 

\subsection*{Analyses of experimental data}
\ We first analyzed the $F=0$ experimental data for Kin1~\cite{Walter_2012,nishiyama2002chemomechanical} in order to obtain the eight parameters at zero load by fitting Eq.(\ref{eq:run_dist}) to the run length distribution, with the constraint that the average velocity, $J^+ - J^- = 132.8\ \steps/s$ at [T] = 1mM \cite{Walter_2012}, and the ratio of forward over backward steps $J^+/J^- = 221$ at [T] = 10$\uMol$ and [T] = 1mM ~\cite{nishiyama2002chemomechanical}. 
We also used the load dependence of the average velocity at $1\mMol$ and $10\uMol$ ATP concentration in 
Ref.~\citenum{nishiyama2002chemomechanical} to obtain the parameters that depend on $F$ and [T]. 
Following previous studies, we set $F_d=3$ pN~\cite{vu2016discrete,M_ller_2010} and $|d^+|+|d^-|=2.9$ nm~\cite{nishiyama2002chemomechanical}. 
Overall we chose the fitting parameters to be $k_0$, $K_T$, $k_0^+$, and $d^+$ out of the eight parameters in our model. 
The best fit parameters are listed in Table~\ref{Table:param_Hancock} and Table~\ref{Table:param_Tomishige} for 2HB and 1HB model, respectively. 
It is worth pointing out that  $k_0^+$ and $k_0^-$ for both the 1HB and 2HB models are fairly close to each other, and are in rough accord with our previous study that did not consider [T]-dependence~\cite{vu2016discrete}. 
Similarly, the distances to the transition state when $F \ne 0$ ($d^+$ and $d^-$) for both the schemes are not that dissimilar (Table~\ref{Table:param_Hancock} and Table~\ref{Table:param_Tomishige}). 

\begin{table}[]
\centering
\caption{\label{Table:param_Hancock} Extracted parameters for 2HB model.}
\begin{tabular}{lllll}
\midrule
$k_0$      &$787.0(s^{-1})$  &\ \ \ &$d^+$&$1.6~(\text{nm})$\\
$k_0^+$    &$185.5(s^{-1})$     &\ \ \  &$d^-$&$1.3(\text{nm})$ \\
$k_0^-$    &$0.8(s^{-1})$&\ \ \ &$F_d$&$3.0(\text{pN})$\\
$\gamma_0$ &$2.4(s^{-1})$&\ \ \ & $K_T$ &$594.0(\mu \text{M})$ \\ 
\bottomrule
\end{tabular}
\end{table}
\begin{table}[]
\centering
\caption{\label{Table:param_Tomishige} Extracted parameters for 1HB model.}
\begin{tabular}{lllll}
\midrule
$k_0$      &$538.0(s^{-1})$  &\ \ \ &$d^+$&$1.9~(\text{nm})$\\
$k_0^+$    &$184(s^{-1})$     &\ \ \  &$d^-$&$1.0(\text{nm})$ \\
$k_0^-$    &$0.8(s^{-1})$&\ \ \ &$F_d$&$3.0(\text{pN})$\\
$\gamma_0$ &$3.0(s^{-1})$&\ \ \ & $K_T$ &$21.0(\mu \text{M})$ \\  
\bottomrule
\end{tabular}
\end{table}

\begin{figure}[]
\centering
\includegraphics[width=0.35\textwidth]{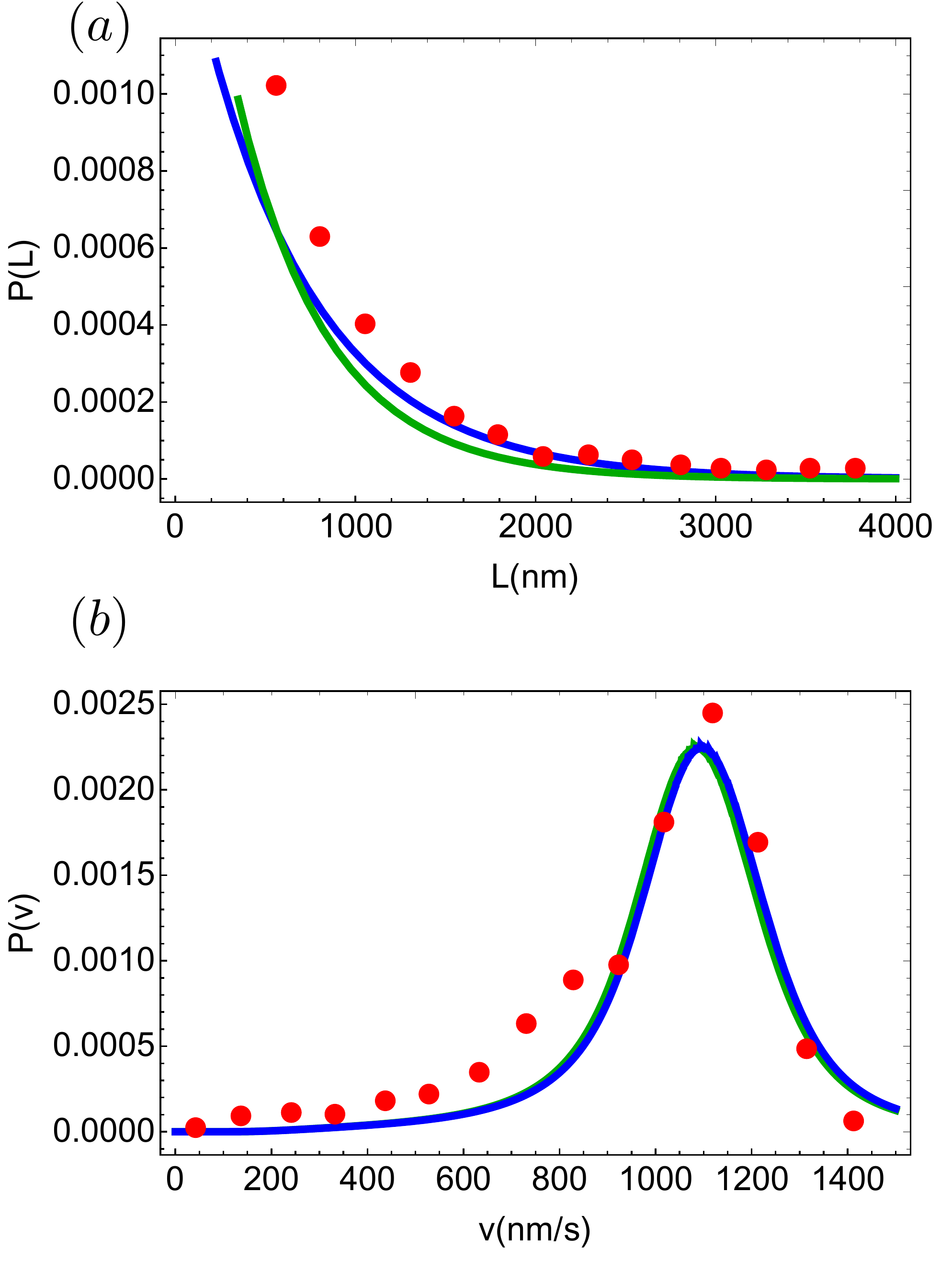}
\caption{\label{Dist_fit_Hancock} 
Simultaneous fits of $P(L)$ ($L= sn$ with $s$ = 8.2 nm) and $P(v)$ at zero load for Kin1 to the experimental data given in \cite{Walter_2012}. Red circles are  from experiment and the blue and green lines are results from our theory, for the 2HB and 1HB model, respectively.  (a) Run length distribution.   
(b) Velocity distribution of Kin1. The comparison shows not only that the theory reproduces the measured data well but the overlap of the blue and green lines shows that at zero load the differences in ATP waiting states are not reflected in the distributions of the run length and velocity.  
}
\end{figure}

\begin{figure}[t]
\centering
\includegraphics[width=0.33\textwidth]{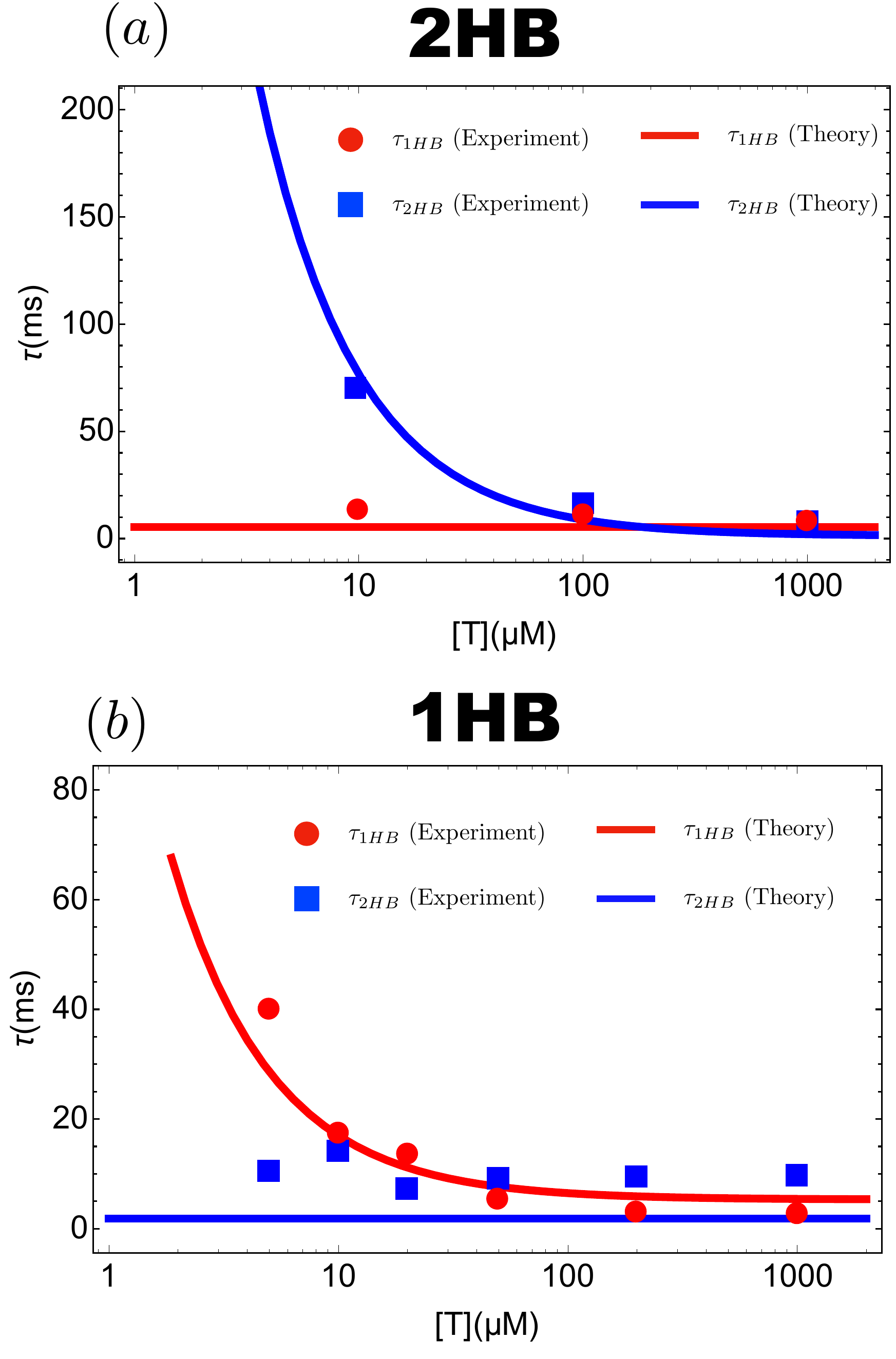}
\caption{\label{Fig:tau1HB_2HB}Mean dwell time for the 2HB state ($\tau_{2HB}$) and 1HB state ($\tau_{1HB}$) as a function of ATP concentration at $F=0$. The upper panel is for the 2HB model [Fig.\ref{fig1}(c)] and the lower panel is for the 1HB model [Fig.\ref{fig1}(d)]. Red circles and blue squares are taken from the experiment by Mickolajczyk {\it et al.}~\cite{Mickolajczyk_2015} (upper panel) and Isojima {\it et al.}~\cite{isojima2016direct} (lower panel). Lines are the theoretical predictions for the dwell times for 1HB state and 2HB state. Note that in (a) $\tau_{1HB}$ is [T] independent whereas in (b) $\tau_{2HB}$ does not depend on [T].
}
\end{figure}

In order to ascertain that our kinetic schemes for the 2HB and 1HB model provide a faithful description of the data of Mickolajczyk {\it et al.}~\cite{Mickolajczyk_2015} and Isojima {\it et al.}~\cite{isojima2016direct}, we compare the life-time of the 1HB [$\tau_{1HB} = 1/(\kp+\km+\gamma)$] and 2HB ($\tau_{2HB} = 1/k$) with the experimental measurements.
As shown in Fig.~\ref{Fig:tau1HB_2HB} the agreement for both the scenarios is excellent, indicating that our theory captures the results of the experiments~\cite{Mickolajczyk_2015,isojima2016direct} accurately.
We hasten to emphasize that the data from Mickolajczyk {\it et al.} and Isojima {\it et al.} were not used for fitting.  The agreement is a genuine emergent feature of our kinetic model, which lends credence to the additional predictions made below.

\subsection*{Velocity distribution is bimodal when $\boldsymbol{F \ne 0}$}
\begin{figure}[]
\centering
\includegraphics[width=0.35\textwidth]{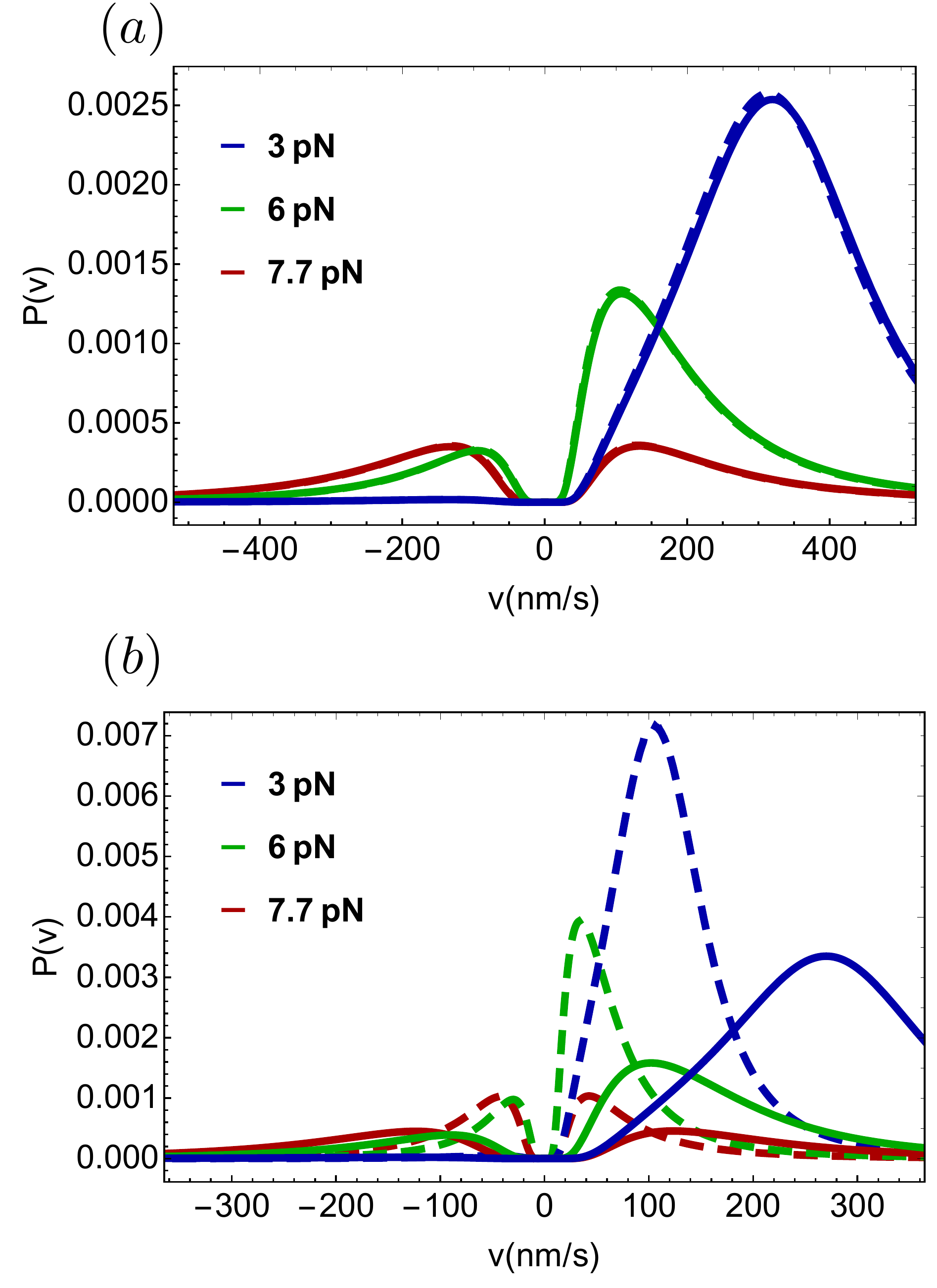}
\caption{\label{Vel_dist_Tomishige_Hancock} 
Velocity distributions for $v\neq0$ predicted by our theory for different loads and ATP concentrations. Lines are for the 2HB model [Fig.\ref{fig1}(c)] and dashed lines are for the 1HB model [Fig.\ref{fig1}(d)]. Colors represent different load applied to kinesin. (a) Velocity distribution at 1mM ATP concentration. 
(b) Velocity distribution for 10$\mu$M ATP concentration.   
}
\end{figure}

\ We use the analytical solutions for $P(n)$ [Eq.~(\ref{eq:run_dist})] and $P(v)$ [Eq.~(\ref{eq:pv})] in order to predict how the distributions of run length and velocity change over a broad range of load and ATP concentrations for the two models (see Fig.~\ref{Vel_dist_Tomishige_Hancock}).
First, we note that the bimodality of the velocity distribution, originally predicted by Vu {\it et al.}~\cite{vu2016discrete}, is evident at both high ($1\mMol$) and low ($10\uMol$) ATP concentrations.
The peak at the negative $v$ increases as $F$ approaches $F_S$.
As the ATP concentration is lowered the motor slows down and the location of the peak of the velocity distribution becomes closer to zero. 
Second, the $P(v)$s at all values of $F$ when [T] is 1mM are similar in the  1HB and 2HB scenarios (upper panel in Fig.~\ref{Vel_dist_Tomishige_Hancock}), and hence cannot be used to easily distinguish between them when the [T] is high. 
Although the shape of $P(v)$ does depend on the ATP waiting state at low [T] (right panel in Fig.~\ref{Vel_dist_Tomishige_Hancock}), which in principle amenable to experimental test, the small qualitative difference may not be sufficient to discriminate between the waiting states in practice.
To summarize, we showed that the bimodality of $P(v)$ is robust to changes in the concentration of ATP and model used for the ATP waiting states.
This provides experimental flexibility in testing the predicted bimodality. Although the prediction of bimodal behavior as a function of [T] and $F$ is most interesting in its own right, it may be challenging to use $P(v)$ as a probe to determine the nature of the ATP waiting state in conventional kinesin.

\subsection*{Randomness parameters are qualitatively different in the 1HB and 2HB waiting states for ATP}
\ Fluctuation analyses in molecular motors are performed using the so-called chemical and mechanical randomness parameters~ \cite{schnitzer1997kinesin,visscher1999single,taniguchi2008forward}. 
The former describes the fluctuation of the enzymatic states of the motor, and is given by $r_{C}=(\langle \tau^2 \rangle-\langle \tau \rangle^2) /\langle \tau \rangle^2$\normalsize. 
Here, $\tau$ is the dwell time of the motor at one site and the bracket denotes average over an ensemble of motors. 
The mechanical randomness parameter is given by, \small$r_{M}=\lim_{t \to \infty}(\langle n^2(t)\rangle  -\langle  n(t)\rangle^2)/\langle n(t)\rangle$\normalsize. 
It can be shown that $r_{C}=r_{M}$ and bounded from 0 to 1 if there are no backward steps~\cite{schnitzer1995statistical}. 
However, it is possible that $r_{M}$ increases beyond 1 when load acts on the motor due to the presence of backward steps. 
We found analytical expressions for $r_{C}$ and $r_{M}$, which allowed us to compare the deviation of the two kinds of randomness parameter as the external load increases. 
We can recover $r_{C}$ from $r_{M}$ by using the relation \small $r_{C}=[(2P_+-1)r_{M}-4P_+(1-P_+)]/(2P_+-1)^2$\normalsize, where $P_+$ is the probability of forward stepping. 
We denote the chemical randomness parameter calculated from mechanical randomness parameter given above as $\rchem$ in order to differentiate it from $r_{C}$, which is not easy to measure experimentally~\cite{schnitzer1995statistical}.
The relationship connecting $r_C$ and $r_M$ has been  derived elsewhere~\cite{shaevitz2005statistical,chemla2008exact}.  
In the SI, we provide an alternate method, which connects between the chemical and mechanical randomness parameters. 
The chemical randomness parameter in our model is written as,
\begin{align} 
\begin{split}
\label{eq:rchem}
r_{C}=\frac{k^2+(k^++k^-+\gamma)^2}{(k+k^++k^-+\gamma)^2}.
\end{split}                 
\end{align}

In order to calculate the moments needed to calculate $r_{M}$, we first obtain the re-normalized probability distribution, $\bar{f}(n>0,t)$, for the position of the motor at time $t$ on the track,
\small
\begin{align} 
\begin{split}
\label{fnt}
\bar{f}(n,t)
=&\frac{1}{C}\sum_{l=0}^{\infty}\frac{\gamma\sqrt{\pi}}{(n+2l)!l!}\text{e}^{-\frac{\xi_1+\xi_2}{2}t}t^{n+2l+\frac{1}{2}} \\ & \frac{k^{n+2l+1}(k^+)^{n+l}(k^-)^l}{|\xi_2-\xi_1|^{n+2l+\frac{1}{2}}} I_{n+2l+\frac{1}{2}}\big(\frac{|\xi_2-\xi_1|}{2}t\big).
\end{split}                 
\end{align}
\normalsize
The normalization constant $C$, which accounts for the detachment of motors is obtained by summing over both positive and negative values of $n$ in the above equation (see SI for details). 
By computing the first and second moments of $\bar{f}$ for $n$ at sufficiently long times, we can obtain an expression for the mechanical randomness parameter $r_{M}$. 
Because $r_C$ in Eq.(\ref{eq:rchem}) depends on ATP, which occurs in different steps in the 2HB and 1HB model [Fig.~\ref{fig1}(c) and (d), respectively], the variation of $r_C$  as a function of [T] could be used to assess the likelihood of the two models.  

\begin{figure*}[b]
\centering
\includegraphics[width=13cm]{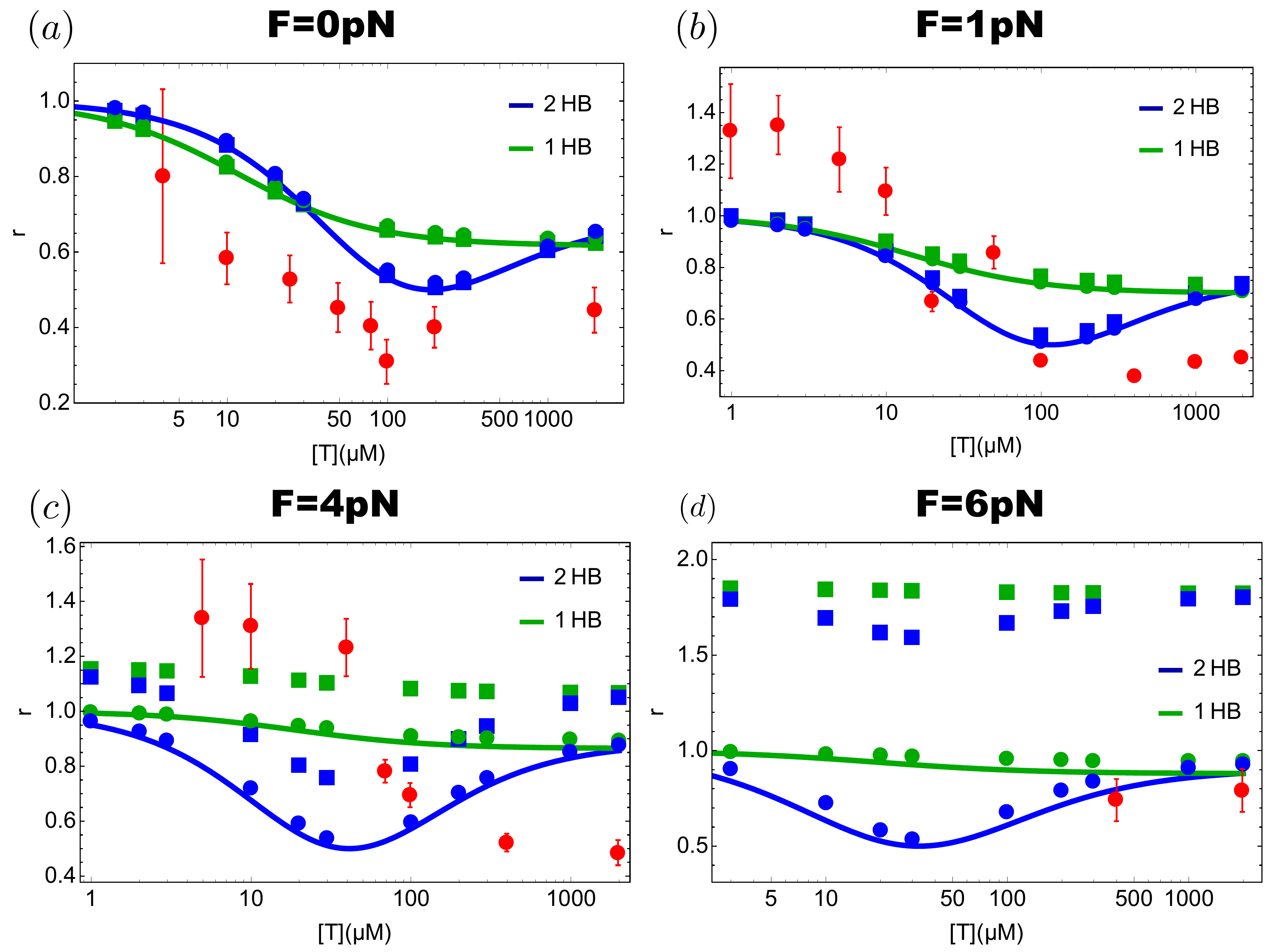}
\caption{\label{fig4} Theoretical prediction of the ATP concentration dependence of the three randomness parameters, $r_{M}$, $r_{C}$, and $\rchem$ at different external loads for the 2HB and 1HB model [Fig.\ref{fig1}(c) and (d), respectively]. Filled squares, filled circles, and lines denote $r_M$, $\rchem$, and $r_C$, respectively. Red circles with error bar in (a) are the experimentally measured randomness parameter at $F=0$ in~\cite{verbrugge2009novel}. Red circles with error bar in (b)-(d) are the randomness parameters measured in \cite{visscher1999single}; (b) for 1.05 pN, (c) for 3.59 pN, and (d) for 5.69 pN. As explained in the discussion section, the values of the randomness parameters in our schemes are always equal or grater than 0.5.  
}
\end{figure*}

In Fig.~\ref{fig4} we plot the randomness parameters, $r_{M}$, $\rchem$, and $r_{C}$ for the kinetic schemes in Fig.~\ref{fig1}(c) and Fig.~\ref{fig1}(d) as a function of ATP concentration at different loads. 
The dependence on ATP concentration of the mechanical randomness parameters for kinesin have been previously reported~\cite{schnitzer1997kinesin,visscher1999single,verbrugge2009novel}. We plotted  the randomness parameter obtained in the experiment by Visscher {\it et al.}~\cite{visscher1999single} and Verbrugge {\it et al.}~\cite{verbrugge2009novel} in Fig.~\ref{fig4} in order to assess if the theory captures the experimental behavior. It is clear that the theory and experiments agree only qualitatively with the trends being very similar. It is known from even more complicated models that it is difficult to calculate with high accuracy the dependence of various randomness parameters on $F$ and [T]~\cite{kolomeisky2003simple,visscher1999single}. 
 Because the randomness parameter measures the inverse of the number of rate-limiting states in the cycle, it is not unreasonable that our model may overestimate the randomness parameter. At higher forces our model for chemical randomness is in near-quantitative agreement, ($F$ = 6pN). In addition, we recover the trend observed in experiments ($F$ = 1pN).
At intermediate values of the forces ($F$ = 4pN) the agreement is less accurate. Thus, we surmise that the agreement between theory and experiments is reasonable so that we can discuss the use of these parameters in deciphering the ATP waiting state of kinesin.

Remarkably, the dependence of the  randomness parameters on [T] and $F$ is dramatically different in the two models for the ATP waiting states.
In the 1HB model, the randomness parameters decreases monotonically. 
In sharp contrast, if ATP binds when both heads are engaged with the MT, we predict a non-monotonic function of [T] with a minimum occurring at near [T]~$\approx100\uMol$. 
This finding suggests an alternative, and perhaps a more straightforward  way, of differentiating between two types of waiting states for ATP. 
If the randomness parameters ($r_M$ and $r_C$) could be measured using the higher resolution single molecule experiments~\cite{isojima2016direct}  as a function of [T] and $F$, then the timing of ATP binding to kinesin could be unambiguously determined. 

It is most interesting that at all values of $F$ the model based on the 2HB waiting state the randomness parameters has a clear minimum as the ATP concentration is changed whereas in the 1HB model the decrease is monotonic and is almost flat as $F$ increases. The difference can be appreciated by noting that in the 2HB model the rate determining step for completing a step changes as [T] is increased from a low value. In particular, at low [T] the rate limiting step is the 2HB$\rightarrow$1HB transition [Fig.~\ref{fig1}(c)] whereas at high [T] the 1HB$\rightarrow$2HB transition is rate limiting [Fig.~\ref{fig1}(c)]. As a consequence of the change in the rate determining step, there is a minimum in the values of the randomness parameter at a critical value of the ATP concentration. Let us write Eq.\ref{eq:rchem} as $r_C = \frac{1 + (x/k)^2}{(1 + x/k)^2}$ where $x = k^++k^-+\gamma$. Using the parameters in Table 1 we determine that at low [T] with $F=0$ ($\frac{x}{k} \gg 1$) whereas at saturating ATP concentration $\frac{x}{k} \approx 0.3$ with a crossover (location of the minimum in the randomness parameter) occurring at $\frac{x}{k} = 1$. The values of [T] at which the randomness parameters are a minimum at different values of $F$ may be estimated using the values in Table 1, which is roughly in accord with the results in Fig.\ref{fig4}. 

In sharp contrast, in the 1HB model the 1HB$\rightarrow$2HB is always slower than the 2HB$\rightarrow$1HB, a feature that is enhanced $F$ increases. This is because the 1HB$\rightarrow$2HB transition is slowed down with load, whereas the 2HB$\rightarrow$1HB is unaffected. In other words, at all values of the ATP concentration the 1HB$\rightarrow$2HB transition is rate limiting with $\frac{x}{k}$ being less than unity.
As a consequence, the chemical randomness parameter is nearly monotonic and is close to unity at all values of $F$, thus making $r_C$ almost independent of [T] (see Fig.\ref{fig4}).

It might be tempting to conclude based on the randomness parameter at zero load reported in~\cite{verbrugge2009novel} [Fig.~\ref{fig4}(a)] that there is a small dip around $100\uMol$ as predicted theoretically using the 2HB model [Fig.\ref{fig1}(c)]. Although not unambiguous, the randomness parameter with external loads measured by Visscher {\it et al.}~\cite{visscher1999single} [Fig.~\ref{fig4}(b)-(d)] apparently shows more or less a monotonic decease with increasing [T], which agrees with the predictions of  the 1HB model [Fig.\ref{fig1}(d)].
We note that  the experiment at zero load [Fig.~\ref{fig4}(a)] was conducted by using fluorescence microscopy and those at non-zero load used  optical trapping technique.
Because of limited temporal resolution in prior experiments, all the measurements of randomness parameter correspond to $r_M$, the mechanical randomness parameter. 
With access to temporal resolution on the order of tens of  microseconds, it may be possible to directly measure the chemical randomness parameter. 
For a fuller understanding of mechano-chemistry of kinesin and in particular how Kin1 waits for ATP, it is desirable to explore the [T] and $F$ dependence of chemical/mechanical parameters using high resolution stepping trajectories.      


\section*{Discussion} 
We have introduced a simple model for stepping of conventional kinesin on microtubule in order to propose  single molecule experiments, which could be used to discriminate between the waiting states for ATP binding to the leading head. We derived analytical solutions for the run length and velocity distributions and various randomness parameters as a function of ATP-concentration and external resistive load. For both the 1HB model and 2HB models $P(n)$ is independent of [T], which is in good agreement with experiments except at very low [T] concentrations, perhaps due to enhanced probability of spontaneous detachment~\cite{verbrugge2009novel,visscher1999single}. 
Therefore, although $P(n)$ could be measured readily it cannot be easily used to distinguish between the two distinct waiting states. The distribution of velocity, which exhibits bimodal behavior at $F \ne 0$, is qualitatively similar both at high and low ATP concentrations. The velocity distribution does differ quantitatively at low ATP concentrations as $F$ is varied [see Fig.~\ref{Vel_dist_Tomishige_Hancock}(b)].  The most significant finding is that that the randomness parameters, which could be measured readily, shows {\it qualitative} differences as a function of $F$ and [T] between the 2HB and 1HB waiting state for ATP. 

{\bf Predicted Bi-modality in the velocity distribution is independent of the ATP waiting  states:}
Since the mean run length does not depend significantly on the ATP concentration for Kin1~\cite{verbrugge2009novel,visscher1999single} it follows that the mean position from which the motor detaches from the MT is roughly the same irrespective of ATP concentrations. Thus, [T] would not affect the spatial resolution needed to observe the predicted bimodality in the velocity distribution. However, since the average velocity of kinesin increases with [T], it  would affect the temporal resolution needed to validate the  shape in $P(v)$. We propose that it would be easier for experimentalists to observe the theoretical prediction that $P(v)$ is bimodal at lower ATP concentrations. This most interesting prediction, made a few years ago~\cite{vu2016discrete} without considering the [T]-dependence in contrast to this study, awaits experimental tests. 

{\bf Randomness parameters are dramatically different between the two waiting states:} 
We predict that the [T] and $F$ dependence of the randomness parameters, which is an  estimate of the minimum number of rate limiting states in kinesin, holds the key in assessing the relevance of the two waiting states. 
Since the theory for both the 2HB and 1HB model consider only two states, the calculated randomness parameters cannot be below 0.5. Therefore, it might be tempting to conclude that our predictions may not be realizable in experiments because it has been advocated that more than two states might be needed to fit the experimental data~\cite{fisher2001simple,liepelt2007kinesin}. However, we argue that the qualitative features of the [T]-dependence of the randomness parameter elucidated using our theory should be observable in experiments using the following reasoning. 
Because kinesin has only one ATP-dependent rate per step and the rest of the rates do not depend on ATP, just as in our model, the change of randomness parameter as a function of [T] is only affected by the step that depends on ATP concentration. 
On the other hand, we compressed many potentially relevant states into one internal state that are unaffected by [T]. 
As a consequence, we expect that when [T] becomes large, our model might overestimate the values of the randomness parameters by a factor that is  proportional to the number of actual ATP-independent internal states.
Indeed, if we shift our values for $r$ in Fig.\ref{fig4}(a) so that they match the experimental values at high [T], we would attain an excellent agreement with the data.
The presence of force might further complicate the interplay between internal states.
Nevertheless, the qualitative difference between the 1HB and 2HB model should be amenable to experimental verification.
Therefore, we believe that accurate measurements of $r_M$ and $r_C$ using high temporal resolution experiments will be most useful in filling a critical missing gap in the catalytic cycle of Kin1. 



{\bf Status of experiments and relation to theory:}
Randomness parameters have been measured previously using fluorescence microscopy~\cite{verbrugge2009novel} and optical trapping~\cite{schnitzer1997kinesin,visscher1999single}. The experimental set up in~\cite{verbrugge2009novel} did not contain cargo whereas the stepping trajectories in the optical trapping experiments were measured by monitoring the time-dependent movement of  a bead attached to the coiled-coil\cite{schnitzer1997kinesin,visscher1999single}. Both experiments from Hancock and coworkers~\cite{Mickolajczyk_2015} and Tomishige and coworkers~\cite{isojima2016direct}, employ innovative experimental methods, which are different from the techniques previously used to measure the randomness parameters. These experiments also did not have cargo but a large AuNP (with diameters between 20 to 40 nm) was attached to different sites on one of the motor heads. The AuNP experiments should have sufficient temporal and spatial resolution to extract both the mechanical and chemical randomness parameters as a function of ATP concentration. The current iSCAT or experiments based on dark field microscopy may not be able to measure the randomness parameter as a function of $F$, which would require attaching a bead (cargo) that would not interfere with the dynamics of AuNP. 
Nevertheless, measurements of randomness parameters using the experimental constructs in~\cite{Mickolajczyk_2015,isojima2016direct} as a function of [T] but with $F=0$ can be made. Such studies are needed to test our predictions (Fig.\ref{fig4}a), which would hopefully  provide insights into the ATP waiting state of kinesin. 

{\bf Mechano-chemistry of the backward step:}
In our model for the 1HB waiting state [Fig.\ref{fig1}(d)], we assumed that the rate of the backward stepping depends on [T] in the same manner as the rate for the forward step. It stands to reason that any step should consume ATP, and consequently $k^-$ should also depend on [T]. Indeed, it has been argued that Kin1 walks backwards by a hand-over-hand mechanism by hydrolyzing ATP in much the same as it does when moving forward~\cite{nishiyama2002chemomechanical,carter2005mechanics}. The observation that the ratio of the probability of taking forward to backward steps as a function of $F$ at two [T] concentrations (1 mM and 10 $\uMol$) superimpose [see Fig.4b in~\cite{nishiyama2002chemomechanical}] lends support to the supposition that $k^{-}$ should also depend on [T]. Our 2HB and 1HB models [Fig.\ref{fig1}(c) and (d), respectively], which consider ATP binding even for backward steps, leads to the prediction that both the run length and the fraction of forward step to backward step are independent of [T], as shown in the experiments~\cite{nishiyama2002chemomechanical,carter2005mechanics}. 
In addition, several of theoretical models have been proposed  to rationalize the [T]-dependence of the backward step~\cite{liepelt2007kinesin,hwang2018energetic,clancy2011universal,hyeon2009kinesin,carter2005mechanics,sumi2017design}. Therefore, our assumption that $k^{-}$ depends on [T] seems justifiable.   

However, the mechanism, especially in structural terms, of the backward step is not fully understood~\cite{liepelt2007kinesin,clancy2011universal,hyeon2009kinesin,carter2005mechanics}. Therefore, it is important to entertain the possibility that $k^-$ has negligible dependence on [T].
Note that the magnitude of $k^-$ is non-negligible only in the presence of substantial load.  At very low forces one could neglect the [T]-dependence of $k^-$. Under these conditions the mechanisms for forward and backward steps need not be the same.

\begin{figure*}[h]
\centering
\includegraphics[width=12.8cm]{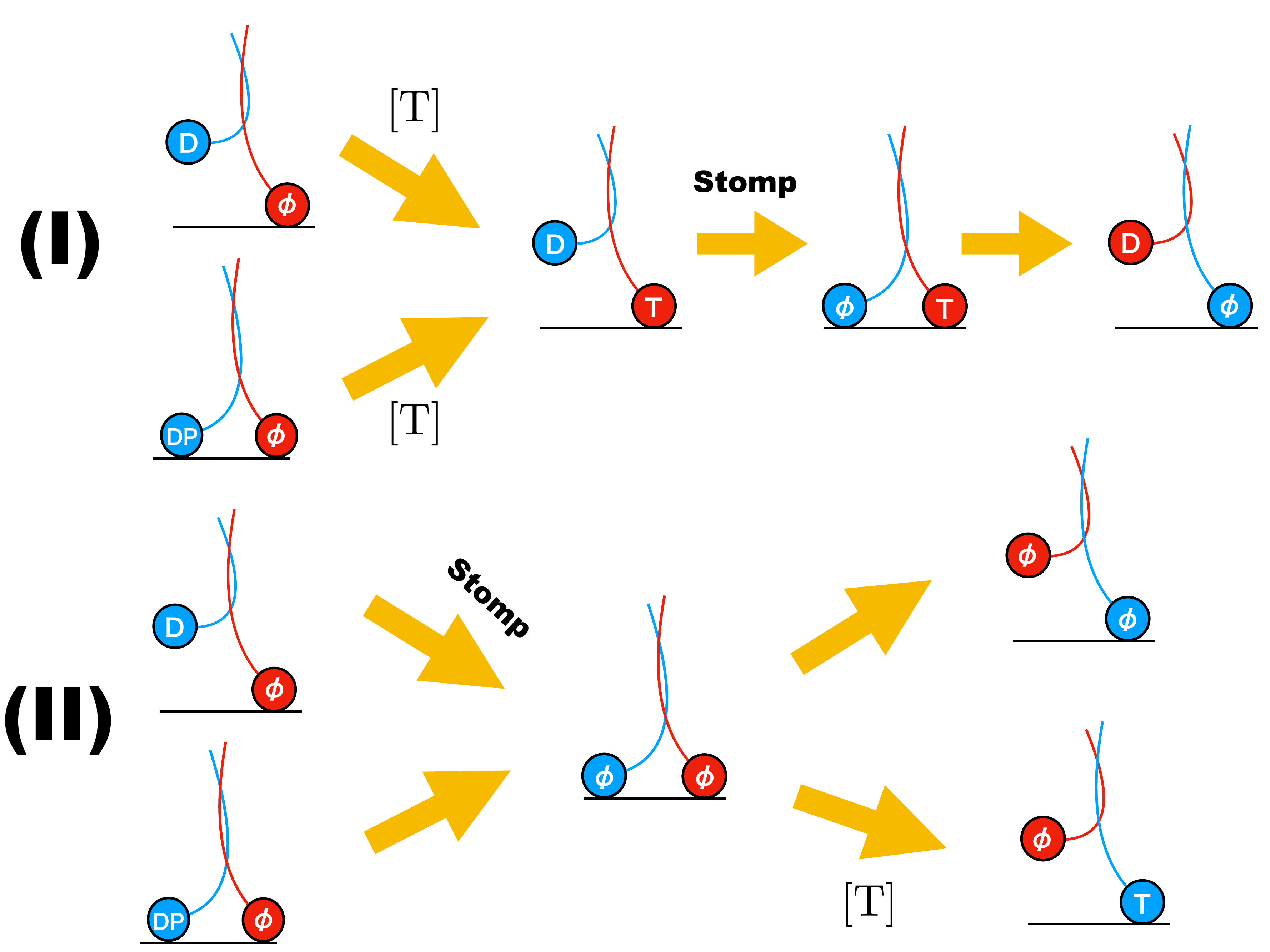}
\caption{\label{backward} Plausible backward step mechanisms for kinesin. Upper panel corresponds to pathway (I) explained in the discussion section. In this case the [T] dependence is identical to forward stepping. Lower panel is pathway (II) in which [T] dependence could be different from the forward stepping. T, D, DP, and $\phi$ stands ATP, ADP, (ADP + phosphate), and no nucleotide state, respectively. The two pathways partially adapted from Ref.\cite{clancy2011universal}. In the upper panel, after ATP binding to the leading head (shown in red) the the neck linker docks. Consequently, the backward step along this pathway would only be possible at higher loads.
}
\end{figure*}

There are at least two possible pathways (see Fig.\ref{backward}) by which Kin1 could take backward steps. 
(I) Let us consider that ATP binds to the LH in either 2HB state or 1HB state and the TH  detaches  with bound ADP. In order for a backward step to occur, the TH has to release ADP and perform a "foot stomp" (return to the starting position). Although to  date there is no evidence for either TH or LH foot stomping in Kin1, they have been observed in Myosin V in the absence of external load~\cite{kodera2010video}. The probability of foot stomping could certainly increases if $F\ne0$, but is improbable in the absence of load. If stomping were to occur, then both the heads would be bound to the MT with the LH containing ATP (third step in pathway I in Fig.\ref{backward}). After TH stomping, ATP should be hydrolyzed and the inorganic phosphate released from the LH, which would lead to backward stepping. 
This pathway results in identical [T] dependence for forward and backward steps. Consequently, the [T] independent characteristics of Kin1, such as $P(n)$, can be explained by this scenario. (II) Let us consider another possibility for backward steps. Before ATP binds to the LH in either 1HB state or 2HB state, ADP is released from TH, leading to 2HB state with both the heads being nucleotide free pathway II in Fig.\ref{backward}). For backward state to occur from this state, the LH should detach from the 2HB state either spontaneously or by binding ATP. The latter event, which would induce neck linker docking, and hence propel the TH forward would tend to suppress the probability of backward steps. If the former were to occur then it might be possible, especially if $F\ne0$, that $k^-$ might not depend on [T]. 

The theory developed based on the scheme in Fig.\ref{fig1}(d) does not account for  the possibility that backward step rate  may not depend on [T]. For completeness, we created in the SI a variant of the 1HB model, corresponding to scenario (II), by setting $k^-$ in Fig.\ref{fig1}(d) to be independent of [T]. The results in the SI show that regardless of the dependence or independence of $k^{-}$ on [T] the qualitative differences in the randomness parameters as a function of $F$ and [T] between the 1HB and 2HB model remain. Thus, the theoretical predictions are robust, suggesting that high temporal resolution experiments that measure randomness could be used to discriminate between the two waiting states for ATP. 
\section*{Conclusion}
It has been challenging to decipher how exactly kinesin waits for ATP to bind to the leading head. Recent experiments have arrived at contradictory conclusions using similar experimental techniques. Although one cannot rule out the possibility that different kinesin constructs and the location of attachment of the gold nanoparticle used in these experiments might lead to different stepping trajectories, it is important to consider the theoretical consequences of the two plausible waiting states of kinesin.  To discriminate between the 1HB and 2HB waiting states,  we developed simple models, allowing us to calculate analytically and fairly accurately a number of measurable quantities. The theory predicts that there ought to be qualitative differences in the randomness parameters as a function of load and ATP concentration. Although the force dependence of the randomness parameters have been previously measured using optical trap techniques, it would be most interesting to repeat these measurements using the constructs used in the most recent experiments~\cite{Mickolajczyk_2015,isojima2016direct}.   In addition, measurements of the load dependence of the randomness parameters using a combination of dark field microscopy methods in combination with optical traps would be most illuminating to verify many of the predictions outlined here.

\subsection*{Materials and Methods}

We created two stochastic kinetic models in order to calculate a number of quantities associated with the stepping kinetics of conventional kinesin. The sketch of the 1HB and 2HB models and the pathways leading from the resting state to the target binding states along with the rates and [T]-dependence are given in Fig.~\ref{fig1}. The model, a generalization of the one introduced previously \cite{vu2016discrete} in order to include the important aspect of [T]-dependence, can be solved exactly, thus allowing us to calculate $P(v)$ and the different randomness parameters for the two different scenarios for the waiting states for ATP binding (see SI appendix for details). Despite the simplicity, we show in the SI that the model does quantitatively reproduce the experimentally measured [T]-dependent force-velocity relation using physically reasonable parameters for the rates describing the two schemes [Fig.~\ref{fig1}(c) and (d)].

\showmatmethods{} 

\acknow{We are grateful to Ahmet Yildiz and William Hancock for their interest and useful comments. This work was supported by National Science Foundation Grant CHE-1900093. Additional support was provided by Collie-Welch Reagents Chair F-0019.}

\showacknow{} 

\bibliography{mybib}

\begin{thebibliography}{10}

\bibitem{Mickolajczyk_2015}
Mickolajczyk KJ, et~al. (2015) Kinetics of nucleotide-dependent structural
  transitions in the kinesin-1 hydrolysis cycle.
\newblock {\em Proceedings of the National Academy of Sciences}
  112(52):E7186--E7193.

\bibitem{isojima2016direct}
Isojima H, Iino R, Niitani Y, Noji H, Tomishige M (2016) Direct observation of
  intermediate states during the stepping motion of kinesin-1.
\newblock {\em Nature chemical biology} 12(4):290.

\bibitem{Svoboda93Nature}
SVOBODA K, SCHMIDT C, SCHNAPP B, BLOCK S ({1993}) {Direct Observation Of
  Kinesin Stepping by BY Optical Trapping Interferometry}.
\newblock {\em {Nature}} {365}({6448}):{721--727}.

\bibitem{asbury2003kinesin}
Asbury CL, Fehr AN, Block SM (2003) Kinesin moves by an asymmetric
  hand-over-hand mechanism.
\newblock {\em Science} 302(5653):2130--2134.

\bibitem{Block07BJ}
Block SM (2007) Kinesin motor mechanics: Binding, stepping, tracking, gating
  and limping.
\newblock {\em Biophys. J.} 92:2986--2995.

\bibitem{mori2007kinesin}
Mori T, Vale RD, Tomishige M (2007) How kinesin waits between steps.
\newblock {\em Nature} 450(7170):750.

\bibitem{yildiz2004kinesin}
Yildiz A, Tomishige M, Vale RD, Selvin PR (2004) Kinesin walks hand-over-hand.
\newblock {\em Science} 303(5658):676--678.

\bibitem{MIKI2005467}
Miki H, Okada Y, Hirokawa N (2005) Analysis of the kinesin superfamily:
  insights into structure and function.
\newblock {\em Trends in Cell Biology} 15(9):467 -- 476.

\bibitem{hackney1995highly}
Hackney DD (1995) Highly processive microtubule-stimulated atp hydrolysis by
  dimeric kinesin head domains.
\newblock {\em Nature} 377(6548):448.

\bibitem{Pilling_2006}
Pilling AD, Horiuchi D, Lively CM, Saxton WM (2006) Kinesin-1 and dynein are
  the primary motors for fast transport of mitochondria in drosophila motor
  axons.
\newblock {\em Molecular Biology of the Cell} 17(4):2057--2068.

\bibitem{yildiz2008intramolecular}
Yildiz A, Tomishige M, Gennerich A, Vale RD (2008) Intramolecular strain
  coordinates kinesin stepping behavior along microtubules.
\newblock {\em Cell} 134(6):1030--1041.

\bibitem{Walter_2012}
Walter WJ, Ber{\'a}nek V, Fischermeier E, Diez S (2012) Tubulin acetylation
  alone does not affect kinesin-1 velocity and run length in vitro.
\newblock {\em PLoS ONE} 7(8):e42218.

\bibitem{visscher1999single}
Visscher K, Schnitzer MJ, Block SM (1999) Single kinesin molecules studied with
  a molecular force clamp.
\newblock {\em Nature} 400(6740):184.

\bibitem{carter2005mechanics}
Carter NJ, Cross R (2005) Mechanics of the kinesin step.
\newblock {\em Nature} 435(7040):308.

\bibitem{vu2016discrete}
Vu HT, Chakrabarti S, Hinczewski M, Thirumalai D (2016) Discrete step sizes of
  molecular motors lead to bimodal non-gaussian velocity distributions under
  force.
\newblock {\em Physical review letters} 117(7):078101.

\bibitem{Zhang_2017}
Zhang Z, Goldtzvik Y, Thirumalai D (2017) Parsing the roles of neck-linker
  docking and tethered head diffusion in the stepping dynamics of kinesin.
\newblock {\em Proceedings of the National Academy of Sciences}
  114(46):E9838--E9845.

\bibitem{ZHANG2012628}
Zhang Z, Thirumalai D (2012) Dissecting the kinematics of the kinesin step.
\newblock {\em Structure} 20(4):628 -- 640.

\bibitem{schnitzer1997kinesin}
Schnitzer MJ, Block SM (1997) Kinesin hydrolyses one atp per 8-nm step.
\newblock {\em Nature} 388(6640):386.

\bibitem{asenjo2006nucleotide}
Asenjo AB, Weinberg Y, Sosa H (2006) Nucleotide binding and hydrolysis induces
  a disorder-order transition in the kinesin neck-linker region.
\newblock {\em Nature structural \& molecular biology} 13(7):648.

\bibitem{fisher2001simple}
Fisher ME, Kolomeisky AB (2001) Simple mechanochemistry describes the dynamics
  of kinesin molecules.
\newblock {\em Proceedings of the National Academy of Sciences}
  98(14):7748--7753.

\bibitem{liepelt2007kinesin}
Liepelt S, Lipowsky R (2007) Kinesin's network of chemomechanical motor cycles.
\newblock {\em Physical review letters} 98(25):258102.

\bibitem{hwang2016quantifying}
Hwang W, Hyeon C (2016) Quantifying the heat dissipation from a molecular
  motor's transport properties in nonequilibrium steady states.
\newblock {\em The journal of physical chemistry letters} 8(1):250--256.

\bibitem{hwang2018energetic}
Hwang W, Hyeon C (2018) Energetic costs, precision, and transport efficiency of
  molecular motors.
\newblock {\em The journal of physical chemistry letters} 9(3):513--520.

\bibitem{sumi2017design}
Sumi T (2017) Design principles governing chemomechanical coupling of kinesin.
\newblock {\em Scientific reports} 7(1):1163.

\bibitem{Wagoner16JPhysChemB}
Wagoner JA, Dill KA ({2016}) {Molecular Motors: Power Strokes Outperform
  Brownian Ratchets}.
\newblock {\em {J. Phys. Chem. B}} {120}({26}):{6327--6336}.

\bibitem{milic2014kinesin}
Milic B, Andreasson JO, Hancock WO, Block SM (2014) Kinesin processivity is
  gated by phosphate release.
\newblock {\em Proceedings of the National Academy of Sciences}
  111(39):14136--14140.

\bibitem{andreasson2015examining}
Andreasson JO, et~al. (2015) Examining kinesin processivity within a general
  gating framework.
\newblock {\em Elife} 4:e07403.

\bibitem{GENNERICH200959}
Gennerich A, Vale RD (2009) Walking the walk: how kinesin and dynein coordinate
  their steps.
\newblock {\em Current Opinion in Cell Biology} 21(1):59 -- 67.
\newblock Cell structure and dynamics.

\bibitem{asenjo2003configuration}
Asenjo AB, Krohn N, Sosa H (2003) Configuration of the two kinesin motor
  domains during atp hydrolysis.
\newblock {\em Nature Structural \& Molecular Biology} 10(10):836.

\bibitem{kawaguchi2001nucleotide}
Kawaguchi K, Ishiwata S (2001) Nucleotide-dependent single-to double-headed
  binding of kinesin.
\newblock {\em Science} 291(5504):667--669.

\bibitem{asenjo2009mobile}
Asenjo AB, Sosa H (2009) A mobile kinesin-head intermediate during the
  atp-waiting state.
\newblock {\em Proceedings of the National Academy of Sciences}
  106(14):5657--5662.

\bibitem{Alhadeff17PNAS}
Alhadeff R, Warshel A ({2017}) {Reexamining the origin of the directionality of
  myosin V}.
\newblock {\em {Proc.Natl. Acad. Sci.}} {114}({39}):{10426--10431}.

\bibitem{Mukherjee17PNAS}
Mukherjee S, Alhadeff R, Warshel A ({2017}) {Simulating the dynamics of the
  mechanochemical cycle of myosin-V}.
\newblock {\em {Proc. Natl. Acad. Sci.}} {114}({9}):{2259--2264}.

\bibitem{Hyeon11BJ}
Hyeon C, Onuchic JN (2011) {A Structural Perspective on the Dynamics of Kinesin
  Motors}.
\newblock {\em Biophys. J.} 101:2749--2759.

\bibitem{Hyeon07PNAS}
Hyeon C, Onuchic JN (2007) Internal strain regulates the nucleotide binding
  site of the kinesin leading head.
\newblock {\em Proc. Natl. Acad. Sci. U. S. A.} 104:2175--2180.

\bibitem{Hyeon07PNAS2}
Hyeon C, Onuchic JN (2007) Mechanical control of the directional stepping
  dynamics of the kinesin motor.
\newblock {\em Proc. Natl. Acad. Sci. U. S. A.} 104:17382--17387.

\bibitem{sindelar2017tracking}
Sindelar CV, Liu D (2017) Tracking down kinesin's achilles heel with balls of
  gold.
\newblock {\em Biophysical journal} 112(12):2454--2456.

\bibitem{hillfree}
Hill T (1989) {\em Free Energy Transduction and Biochemical Cycle Kinetics}.
\newblock (Springer).

\bibitem{Hill2879}
Hill TL (1988) Interrelations between random walks on diagrams (graphs) with
  and without cycles.
\newblock {\em Proceedings of the National Academy of Sciences}
  85(9):2879--2883.

\bibitem{zhang2017theoretical}
Zhang Y, Kolomeisky AB (2017) Theoretical investigation of distributions of run
  lengths for biological molecular motors.
\newblock {\em The Journal of Physical Chemistry B} 122(13):3272--3279.

\bibitem{nishiyama2002chemomechanical}
Nishiyama M, Higuchi H, Yanagida T (2002) Chemomechanical coupling of the
  forward and backward steps of single kinesin molecules.
\newblock {\em Nature Cell Biology} 4(10):790.

\bibitem{M_ller_2010}
M{\"u}ller MJI, Berger F, Klumpp S, Lipowsky R (2010) Cargo transport by teams
  of molecular motors: Basic mechanisms for intracellular drug delivery.
\newblock {\em Organelle-Specific Pharmaceutical Nanotechnology} pp. 289--309.

\bibitem{taniguchi2008forward}
Taniguchi Y, Yanagida T (2008) The forward and backward stepping processes of
  kinesin are gated by atp binding.
\newblock {\em Biophysics} 4:11--18.

\bibitem{schnitzer1995statistical}
Schnitzer MJ, Block S (1995) Statistical kinetics of processive enzymes in {\em
  Cold spring harbor symposia on quantitative biology}.
\newblock (Cold Spring Harbor Laboratory Press), Vol.{}~60, pp. 793--802.

\bibitem{shaevitz2005statistical}
Shaevitz JW, Block SM, Schnitzer MJ (2005) Statistical kinetics of
  macromolecular dynamics.
\newblock {\em Biophysical journal} 89(4):2277--2285.

\bibitem{chemla2008exact}
Chemla YR, Moffitt JR, Bustamante C (2008) Exact solutions for kinetic models
  of macromolecular dynamics.
\newblock {\em The Journal of Physical Chemistry B} 112(19):6025--6044.

\bibitem{verbrugge2009novel}
Verbrugge S, Van~den Wildenberg SM, Peterman EJ (2009) Novel ways to determine
  kinesin-1's run length and randomness using fluorescence microscopy.
\newblock {\em Biophysical journal} 97(8):2287--2294.

\bibitem{kolomeisky2003simple}
Kolomeisky AB, Fisher ME (2003) A simple kinetic model describes the
  processivity of myosin-v.
\newblock {\em Biophysical journal} 84(3):1642--1650.

\bibitem{clancy2011universal}
Clancy BE, Behnke-Parks WM, Andreasson JO, Rosenfeld SS, Block SM (2011) A
  universal pathway for kinesin stepping.
\newblock {\em Nature structural \& molecular biology} 18(9):1020.

\bibitem{hyeon2009kinesin}
Hyeon C, Klumpp S, Onuchic JN (2009) Kinesin's backsteps under mechanical load.
\newblock {\em Physical Chemistry Chemical Physics} 11(24):4899--4910.

\bibitem{kodera2010video}
Kodera N, Yamamoto D, Ishikawa R, Ando T (2010) Video imaging of walking myosin
  v by high-speed atomic force microscopy.
\newblock {\em Nature} 468(7320):72.

\end{thebibliography}


\begin{thebibliography}{15}%
\makeatletter
\providecommand \@ifxundefined [1]{%
 \@ifx{#1\undefined}
}%
\providecommand \@ifnum [1]{%
 \ifnum #1\expandafter \@firstoftwo
 \else \expandafter \@secondoftwo
 \fi
}%
\providecommand \@ifx [1]{%
 \ifx #1\expandafter \@firstoftwo
 \else \expandafter \@secondoftwo
 \fi
}%
\providecommand \natexlab [1]{#1}%
\providecommand \enquote  [1]{``#1''}%
\providecommand \bibnamefont  [1]{#1}%
\providecommand \bibfnamefont [1]{#1}%
\providecommand \citenamefont [1]{#1}%
\providecommand \href@noop [0]{\@secondoftwo}%
\providecommand \href [0]{\begingroup \@sanitize@url \@href}%
\providecommand \@href[1]{\@@startlink{#1}\@@href}%
\providecommand \@@href[1]{\endgroup#1\@@endlink}%
\providecommand \@sanitize@url [0]{\catcode `\\12\catcode `\$12\catcode
  `\&12\catcode `\#12\catcode `\^12\catcode `\_12\catcode `\%12\relax}%
\providecommand \@@startlink[1]{}%
\providecommand \@@endlink[0]{}%
\providecommand \url  [0]{\begingroup\@sanitize@url \@url }%
\providecommand \@url [1]{\endgroup\@href {#1}{\urlprefix }}%
\providecommand \urlprefix  [0]{URL }%
\providecommand \Eprint [0]{\href }%
\providecommand \doibase [0]{http://dx.doi.org/}%
\providecommand \selectlanguage [0]{\@gobble}%
\providecommand \bibinfo  [0]{\@secondoftwo}%
\providecommand \bibfield  [0]{\@secondoftwo}%
\providecommand \translation [1]{[#1]}%
\providecommand \BibitemOpen [0]{}%
\providecommand \bibitemStop [0]{}%
\providecommand \bibitemNoStop [0]{.\EOS\space}%
\providecommand \EOS [0]{\spacefactor3000\relax}%
\providecommand \BibitemShut  [1]{\csname bibitem#1\endcsname}%
\let\auto@bib@innerbib\@empty
\bibitem [{\citenamefont {Abramowitz}\ and\ \citenamefont
  {Stegun}(1964)}]{Handbook_math}%
  \BibitemOpen
  \bibfield  {author} {\bibinfo {author} {\bibfnamefont {M.}~\bibnamefont
  {Abramowitz}}\ and\ \bibinfo {author} {\bibfnamefont {I.~A.}\ \bibnamefont
  {Stegun}},\ }\href@noop {} {\emph {\bibinfo {title} {Handbook of Mathematical
  Functions with Formulas, Graphs, and Mathematical Tables}}}\ (\bibinfo
  {publisher} {Dover, New York},\ \bibinfo {year} {1964})\BibitemShut {NoStop}%
\bibitem [{\citenamefont {Verbrugge}\ \emph {et~al.}(2009)\citenamefont
  {Verbrugge}, \citenamefont {Van~den Wildenberg},\ and\ \citenamefont
  {Peterman}}]{verbrugge2009novel}%
  \BibitemOpen
  \bibfield  {author} {\bibinfo {author} {\bibfnamefont {S.}~\bibnamefont
  {Verbrugge}}, \bibinfo {author} {\bibfnamefont {S.~M.}\ \bibnamefont {Van~den
  Wildenberg}}, \ and\ \bibinfo {author} {\bibfnamefont {E.~J.}\ \bibnamefont
  {Peterman}},\ }\href@noop {} {\bibfield  {journal} {\bibinfo  {journal}
  {Biophysical journal}\ }\textbf {\bibinfo {volume} {97}},\ \bibinfo {pages}
  {2287} (\bibinfo {year} {2009})}\BibitemShut {NoStop}%
\bibitem [{\citenamefont {Visscher}\ \emph {et~al.}(1999)\citenamefont
  {Visscher}, \citenamefont {Schnitzer},\ and\ \citenamefont
  {Block}}]{visscher1999single}%
  \BibitemOpen
  \bibfield  {author} {\bibinfo {author} {\bibfnamefont {K.}~\bibnamefont
  {Visscher}}, \bibinfo {author} {\bibfnamefont {M.~J.}\ \bibnamefont
  {Schnitzer}}, \ and\ \bibinfo {author} {\bibfnamefont {S.~M.}\ \bibnamefont
  {Block}},\ }\href@noop {} {\bibfield  {journal} {\bibinfo  {journal}
  {Nature}\ }\textbf {\bibinfo {volume} {400}},\ \bibinfo {pages} {184}
  (\bibinfo {year} {1999})}\BibitemShut {NoStop}%
\bibitem [{NIS(2010)}]{NISTHandbook}%
  \BibitemOpen
  \href@noop {} {\emph {\bibinfo {title} {NIST Handbook of Mathematical
  Functions Hardback and CD-ROM}}}\ (\bibinfo  {publisher} {Cambridge
  University Press},\ \bibinfo {year} {2010})\BibitemShut {NoStop}%
\bibitem [{\citenamefont {Svoboda}\ \emph {et~al.}(1994)\citenamefont
  {Svoboda}, \citenamefont {Mitra},\ and\ \citenamefont
  {Block}}]{svoboda1994fluctuation}%
  \BibitemOpen
  \bibfield  {author} {\bibinfo {author} {\bibfnamefont {K.}~\bibnamefont
  {Svoboda}}, \bibinfo {author} {\bibfnamefont {P.~P.}\ \bibnamefont {Mitra}},
  \ and\ \bibinfo {author} {\bibfnamefont {S.~M.}\ \bibnamefont {Block}},\
  }\href@noop {} {\bibfield  {journal} {\bibinfo  {journal} {Proceedings of the
  National Academy of Sciences}\ }\textbf {\bibinfo {volume} {91}},\ \bibinfo
  {pages} {11782} (\bibinfo {year} {1994})}\BibitemShut {NoStop}%
\bibitem [{\citenamefont {Schnitzer}\ and\ \citenamefont
  {Block}(1995)}]{schnitzer1995statistical}%
  \BibitemOpen
  \bibfield  {author} {\bibinfo {author} {\bibfnamefont {M.~J.}\ \bibnamefont
  {Schnitzer}}\ and\ \bibinfo {author} {\bibfnamefont {S.}~\bibnamefont
  {Block}},\ }in\ \href@noop {} {\emph {\bibinfo {booktitle} {Cold spring
  harbor symposia on quantitative biology}}},\ Vol.~\bibinfo {volume} {60}\
  (\bibinfo {organization} {Cold Spring Harbor Laboratory Press},\ \bibinfo
  {year} {1995})\ pp.\ \bibinfo {pages} {793--802}\BibitemShut {NoStop}%
\bibitem [{\citenamefont {Shaevitz}\ \emph {et~al.}(2005)\citenamefont
  {Shaevitz}, \citenamefont {Block},\ and\ \citenamefont
  {Schnitzer}}]{shaevitz2005statistical}%
  \BibitemOpen
  \bibfield  {author} {\bibinfo {author} {\bibfnamefont {J.~W.}\ \bibnamefont
  {Shaevitz}}, \bibinfo {author} {\bibfnamefont {S.~M.}\ \bibnamefont {Block}},
  \ and\ \bibinfo {author} {\bibfnamefont {M.~J.}\ \bibnamefont {Schnitzer}},\
  }\href@noop {} {\bibfield  {journal} {\bibinfo  {journal} {Biophysical
  journal}\ }\textbf {\bibinfo {volume} {89}},\ \bibinfo {pages} {2277}
  (\bibinfo {year} {2005})}\BibitemShut {NoStop}%
\bibitem [{\citenamefont {Chemla}\ \emph {et~al.}(2008)\citenamefont {Chemla},
  \citenamefont {Moffitt},\ and\ \citenamefont {Bustamante}}]{chemla2008exact}%
  \BibitemOpen
  \bibfield  {author} {\bibinfo {author} {\bibfnamefont {Y.~R.}\ \bibnamefont
  {Chemla}}, \bibinfo {author} {\bibfnamefont {J.~R.}\ \bibnamefont {Moffitt}},
  \ and\ \bibinfo {author} {\bibfnamefont {C.}~\bibnamefont {Bustamante}},\
  }\href@noop {} {\bibfield  {journal} {\bibinfo  {journal} {The Journal of
  Physical Chemistry B}\ }\textbf {\bibinfo {volume} {112}},\ \bibinfo {pages}
  {6025} (\bibinfo {year} {2008})}\BibitemShut {NoStop}%
\bibitem [{\citenamefont {Hill}(1989)}]{hillfree}%
  \BibitemOpen
  \bibfield  {author} {\bibinfo {author} {\bibfnamefont {T.}~\bibnamefont
  {Hill}},\ }\href@noop {} {\emph {\bibinfo {title} {Free Energy Transduction
  and Biochemical Cycle Kinetics}}}\ (\bibinfo  {publisher} {Springer},\
  \bibinfo {year} {1989})\BibitemShut {NoStop}%
\bibitem [{\citenamefont {Hill}(1988)}]{Hill2879}%
  \BibitemOpen
  \bibfield  {author} {\bibinfo {author} {\bibfnamefont {T.~L.}\ \bibnamefont
  {Hill}},\ }\href@noop {} {\bibfield  {journal} {\bibinfo  {journal}
  {Proceedings of the National Academy of Sciences}\ }\textbf {\bibinfo
  {volume} {85}},\ \bibinfo {pages} {2879} (\bibinfo {year}
  {1988})}\BibitemShut {NoStop}%
\bibitem [{\citenamefont {Nishiyama}\ \emph {et~al.}(2002)\citenamefont
  {Nishiyama}, \citenamefont {Higuchi},\ and\ \citenamefont
  {Yanagida}}]{nishiyama2002chemomechanical}%
  \BibitemOpen
  \bibfield  {author} {\bibinfo {author} {\bibfnamefont {M.}~\bibnamefont
  {Nishiyama}}, \bibinfo {author} {\bibfnamefont {H.}~\bibnamefont {Higuchi}},
  \ and\ \bibinfo {author} {\bibfnamefont {T.}~\bibnamefont {Yanagida}},\
  }\href@noop {} {\bibfield  {journal} {\bibinfo  {journal} {Nature Cell
  Biology}\ }\textbf {\bibinfo {volume} {4}},\ \bibinfo {pages} {790} (\bibinfo
  {year} {2002})}\BibitemShut {NoStop}%
\bibitem [{\citenamefont {Carter}\ and\ \citenamefont
  {Cross}(2005)}]{carter2005mechanics}%
  \BibitemOpen
  \bibfield  {author} {\bibinfo {author} {\bibfnamefont {N.~J.}\ \bibnamefont
  {Carter}}\ and\ \bibinfo {author} {\bibfnamefont {R.}~\bibnamefont {Cross}},\
  }\href@noop {} {\bibfield  {journal} {\bibinfo  {journal} {Nature}\ }\textbf
  {\bibinfo {volume} {435}},\ \bibinfo {pages} {308} (\bibinfo {year}
  {2005})}\BibitemShut {NoStop}%
\bibitem [{\citenamefont {Walter}\ \emph {et~al.}(2012)\citenamefont {Walter},
  \citenamefont {Ber{\'a}nek}, \citenamefont {Fischermeier},\ and\
  \citenamefont {Diez}}]{Walter_2012}%
  \BibitemOpen
  \bibfield  {author} {\bibinfo {author} {\bibfnamefont {W.~J.}\ \bibnamefont
  {Walter}}, \bibinfo {author} {\bibfnamefont {V.}~\bibnamefont {Ber{\'a}nek}},
  \bibinfo {author} {\bibfnamefont {E.}~\bibnamefont {Fischermeier}}, \ and\
  \bibinfo {author} {\bibfnamefont {S.}~\bibnamefont {Diez}},\ }\href {\doibase
  10.1371/journal.pone.0042218} {\bibfield  {journal} {\bibinfo  {journal}
  {PLoS ONE}\ }\textbf {\bibinfo {volume} {7}},\ \bibinfo {pages} {e42218}
  (\bibinfo {year} {2012})}\BibitemShut {NoStop}%
\bibitem [{\citenamefont {M{\"u}ller}\ \emph {et~al.}(2010)\citenamefont
  {M{\"u}ller}, \citenamefont {Berger}, \citenamefont {Klumpp},\ and\
  \citenamefont {Lipowsky}}]{M_ller_2010}%
  \BibitemOpen
  \bibfield  {author} {\bibinfo {author} {\bibfnamefont {M.~J.~I.}\
  \bibnamefont {M{\"u}ller}}, \bibinfo {author} {\bibfnamefont
  {F.}~\bibnamefont {Berger}}, \bibinfo {author} {\bibfnamefont
  {S.}~\bibnamefont {Klumpp}}, \ and\ \bibinfo {author} {\bibfnamefont
  {R.}~\bibnamefont {Lipowsky}},\ }\href {\doibase 10.1002/9780470875780.ch16}
  {\bibfield  {journal} {\bibinfo  {journal} {Organelle-Specific Pharmaceutical
  Nanotechnology}\ ,\ \bibinfo {pages} {289}} (\bibinfo {year}
  {2010})}\BibitemShut {NoStop}%
\bibitem [{\citenamefont {Vu}\ \emph {et~al.}(2016)\citenamefont {Vu},
  \citenamefont {Chakrabarti}, \citenamefont {Hinczewski},\ and\ \citenamefont
  {Thirumalai}}]{vu2016discrete}%
  \BibitemOpen
  \bibfield  {author} {\bibinfo {author} {\bibfnamefont {H.~T.}\ \bibnamefont
  {Vu}}, \bibinfo {author} {\bibfnamefont {S.}~\bibnamefont {Chakrabarti}},
  \bibinfo {author} {\bibfnamefont {M.}~\bibnamefont {Hinczewski}}, \ and\
  \bibinfo {author} {\bibfnamefont {D.}~\bibnamefont {Thirumalai}},\
  }\href@noop {} {\bibfield  {journal} {\bibinfo  {journal} {Physical review
  letters}\ }\textbf {\bibinfo {volume} {117}},\ \bibinfo {pages} {078101}
  (\bibinfo {year} {2016})}\BibitemShut {NoStop}%
\end{thebibliography}%

\end{document}



\title{Supplementary Information \\How kinesin waits for ATP affects the nucleotide and load dependence of the stepping kinetics}


\author{R. Takaki}
\affiliation{Department of Physics, The University of Texas at Austin, Austin, TX 78712}
\author{M. L. Mugnai}
\affiliation{Department of Chemistry, The University of Texas at Austin, Austin, TX 78712}
\author{Y. Goldtzvik}
\affiliation{Department of Chemistry, The University of Texas at Austin, Austin, TX 78712}
\author{D. Thirumalai}
\affiliation{Department of Chemistry, The University of Texas at Austin, Austin, TX 78712}


\date{\today}



\maketitle

\tableofcontents
\newpage

\section{Derivation of the run length distribution}
The summation in Eq.(4) in the main text,
\begin{align} 
\begin{split}
\label{eq:Pn}
P(n)=\sum_{m,l=0}^{\infty}\frac{(m+l)!}{m!l!}\Big(\frac{J^+}{J_T}\Big)^m\Big(\frac{J^-}{J_T}\Big)^l\Big(\frac{J^\gamma}{J_T}\Big)\delta_{m-l,n},
\end{split}                 
\end{align}
can be carried out for $n>0$, leading to, 
\begin{align} 
\begin{split}
P(n>0)&=\Big(\frac{J^+}{J_T}\Big)^n\Big(\frac{J^\gamma}{J_T}\Big)\sum_{l=0}^{\infty}\Big(\frac{J^+J^-}{J_T^2}\Big)^l\frac{(2l+n)}{(n+l)!l!}\\
&=\Big(\frac{J^+}{J_T}\Big)^n\Big(\frac{J^\gamma}{J_T}\Big)\ _2\text{F}_1\Big(\frac{1+n}{2},\frac{2+n}{2};1+n;4\frac{J^+J^-}{J_T^2}\Big),
\end{split}                 
\end{align}
where $\ _2\text{F}_1$ is Gaussian hypergeometric function. 
By using the special case of $\ _2\text{F}_1$ (page 556 of \cite{Handbook_math}), we obtain the following expression for the run length distribution,
\begin{align} 
\begin{split}
\label{eq:run_dist}
P(n\gtrless0)=\Big(\frac{2J^\pm}{J_T+\sqrt{J_T^2-4J^+J^-}}\Big)^{|n|}\frac{J^\gamma}{\sqrt{J_T^2-4J^+J^-}}.
\end{split}                 
\end{align}
In the above equation  $n$ is non-dimensional quantity which denotes the position of the motor on the track. 
Notice that Eq.(\ref{eq:run_dist}) is independent of ATP concentration, which is in accord with experiments~\cite{verbrugge2009novel,visscher1999single} as long as ATP concentration exceeds $\approx 10\mu$M. At lower ATP concentration, the mean run length does moderately depend on [T]. This suggests that at very low [T] the motor head could enter the vulnerable 1HB state spontaneously. In such a state the LH could detach before ATP binds to it. This would occur if the binding time for ATP (inverse of a pseudo first order constant) is greater than the detachment time. 

\section{Derivation of the velocity distribution}
In our model (Fig.1(c) and (d) in the main text), the probability distribution of forward step, backward step, and detachment at time $t$, and the corresponding Laplace transform ($\mathcal{L}(f)=\int_0^\infty \text{e}^{-st}f(t)dt$) are given by the following expressions,
\begin{align} 
\begin{split}
f_+(t)&=\int_{0}^{t}dt^\prime k\text{e}^{-kt^\prime}k^+\text{e}^{-(k^++k^-+\gamma)(t-t^\prime)}\\
&=\frac{kk^+}{k^++k^-+\gamma-k}(\text{e}^{-kt}-\text{e}^{-(k^++k^-+\gamma)t})\\ 
\tilde{f}_+(s)&=\frac{kk^+}{(s+k)(s+k^++k^-+\gamma)};\\ 
\end{split}                 
\end{align}
and
\begin{align} 
\begin{split}
f_-(t)&=\int_{0}^{t}dt^\prime k\text{e}^{-kt^\prime}k^-\text{e}^{-(k^++k^-+\gamma)(t-t^\prime)}\\
&=\frac{kk^-}{k^++k^-+\gamma-k}(\text{e}^{-kt}-\text{e}^{-(k^++k^-+\gamma)t}) \\ 
\tilde{f}_-(s)&=\frac{kk^-}{(s+k)(s+k^++k^-+\gamma)};\\ 
\end{split}                 
\end{align}
\begin{align} 
\begin{split}
f_\gamma(t)&=\int_{0}^{t}dt^\prime k\text{e}^{-kt^\prime}\gamma\text{e}^{-(k^++k^-+\gamma)(t-t^\prime)}\\
&=\frac{k \gamma}{k^++k^-+\gamma-k}(\text{e}^{-kt}-\text{e}^{-(k^++k^-+\gamma)t}) \\ 
\tilde{f}_\gamma(s)&=\frac{k\gamma}{(s+k)(s+k^++k^-+\gamma)}.
\end{split}                 
\end{align}
The probability that the motor takes $m$ forward steps and $l$ backward steps before detaching at time $t$ can be written as,
\begin{align} 
\begin{split}
\label{eq:factorial}
f(m,l,t)=\frac{(m+l)!}{m!l!}\int_{0}^{t}dt_{m+l}\int_{0}^{t_{m+l}}dt_{m+l-1}\cdots \int_{0}^{t_3}dt_2\int_{0}^{t_2}dt_1\\\prod_{i=1}^{m}f_+(t_i-t_{i-1})\prod_{i=m+1}^{m+l}f_-(t_i-t_{i-1})f_\gamma(t-t_{m+l}).
\end{split}                 
\end{align}
The factorial term in Eq.(\ref{eq:factorial}) accounts for the number of ways in which the motor can take $m$ forward steps and $l$ backward steps for a net displacement of $m-l$.
In Laplace space, Eq.(\ref{eq:factorial}) is a multiplication of all the dwell time distributions up to the time of detachment, 
\begin{align} 
\begin{split}
\tilde{f}(m,l,s) =& \frac{(m+l)!}{m!l!}(\tilde{f}_+(s))^{m}(\tilde{f}_-(s))^{l}\tilde{f}_\gamma(s)\\ 
=& \frac{(m+l)!}{m!l!}k^{m+l+1}(k^+)^m(k^-)^l\gamma\frac{1}{(s+k)^{m+l+1}(s+k^++k^-+\gamma)^{m+l+1}}.
\end{split} 
\end{align}
In order to obtain $f(m,l,t)$ by inverse Laplace transform, we need to evaluate the following integral,
\begin{align} 
\begin{split}
\label{cint}
\frac{1}{2\pi i}\int_{c-i\infty}^{c+i\infty}\frac{1}{(s+\xi_1)^{n+1}(s+\xi_2)^{n+1}}\text{e}^{st}ds,
\end{split} 
\end{align}
where we set $m+l=n$, $\xi_1=k$, and $\xi_2=k^++k^-+\gamma$.
This can be done using the standard residue theorem.
For the generic case of $\xi_1\neq\xi_2$, there are two distinct poles of order $n+1$.
\begin{align} 
\begin{split}
\text{Res}\Big[\frac{1}{(s+\xi_1)^{n+1}(s+\xi_2)^{n+1}}\text{e}^{st}\Big]\Big|_{s=-\xi_1}&=\frac{1}{n!}\frac{d^n}{ds^n}\Big[\frac{1}{(s+\xi_2)^{n+1}}\text{e}^{st}\Big]\Big|_{s=-\xi_1} \\ 
&=\frac{1}{n!}\sum_{p=0}^{n}(-1)^p\binom{n}{p}\frac{t^{n-p}\text{e}^{-\xi_1t}}{(-\xi_1+\xi_2)^{n+p+1}}\frac{(n+p)!}{n!},
\end{split} 
\end{align}
\begin{align} 
\begin{split}
\text{Res}\Big[\frac{1}{(s+\xi_1)^{n+1}(s+\xi_2)^{n+1}}\text{e}^{st}\Big]\Big|_{s=-\xi_2}&=\frac{1}{n!}\frac{d^n}{ds^n}\Big[\frac{1}{(s+\xi_1)^{n+1}}\text{e}^{st}\Big]\Big|_{s=-\xi_2} \\ 
&=\frac{1}{n!}\sum_{p=0}^{n}(-1)^p\binom{n}{p}\frac{t^{n-p}\text{e}^{-\xi_2t}}{(-\xi_2+\xi_1)^{n+p+1}}\frac{(n+p)!}{n!}.
\end{split} 
\end{align}
Thus, by combining Eqs.(10.47.9), (10.49.1), and (10.49.12) of the NIST handbook of Mathematics \cite{NISTHandbook}, the inverse Laplace transform in Eq.(\ref{cint}) becomes,
\begin{align} 
\begin{split}
(\ref{cint})&=\frac{1}{n!}\sum_{p=0}^{n}(-1)^p\binom{n}{p}\frac{t^{n-p}\text{e}^{-\xi_1t}}{(-\xi_1+\xi_2)^{n+p+1}}\frac{(n+p)!}{n!}+\frac{1}{n!}\sum_{p=0}^{n}(-1)^p\binom{n}{p}\frac{t^{n-p}\text{e}^{-\xi_2t}}{(-\xi_2+\xi_1)^{n+p+1}}\frac{(n+p)!}{n!} \\ 
&=\frac{t^n}{\sqrt{\pi}n!}\text{e}^{-\frac{\xi_1+\xi_2}{2}t}\frac{\sqrt{(\xi_1-\xi_2)t}}{(\xi_2-\xi_1)^{n+1}}K_{n+\frac{1}{2}}(\frac{\xi_1-\xi_2}{2}t) + \frac{t^n}{\sqrt{\pi}n!}\text{e}^{-\frac{\xi_1+\xi_2}{2}t}\frac{\sqrt{(\xi_2-\xi_1)t}}{(\xi_1-\xi_2)^{n+1}}K_{n+\frac{1}{2}}(\frac{\xi_2-\xi_1}{2}t),
\end{split} 
\end{align}
where $K$ is the modified Bessel function of the second kind.\\
Hence, Eq.(\ref{eq:factorial}) becomes,
\begin{align} 
\begin{split}
\label{eq:fmlt}
f(m,l,t)&=
\frac{\gamma}{m!l!}\frac{1}{\sqrt{\pi}}\text{e}^{-\frac{\xi_1+\xi_2}{2}t}t^{m+l}k^{m+l+1}(k^+)^m(k^-)^l  \Big[\frac{\sqrt{(\xi_1-\xi_2)t}}{(\xi_2-\xi_1)^{m+l+1}}\ K_{m+l+\frac{1}{2}}\big(\frac{\xi_1-\xi_2}{2}t\big)\\
&+ \frac{\sqrt{(\xi_2-\xi_1)t}}{(\xi_1-\xi_2)^{m+l+1}} 
\ K_{m+l+\frac{1}{2}}\big(\frac{\xi_2-\xi_1}{2}t\big)\Big].
\end{split} 
\end{align}
We can simplify the above expression by employing the following identity for modified Bessel function given in Eq.(10.34.2) in \cite{NISTHandbook}, and rewrite it as,
\begin{align} 
\begin{split}
K_{n+\frac{1}{2}}(-x)=-i\big( \pi I_{n+\frac{1}{2}}(x)+(-1)^nK_{n+\frac{1}{2}}(x) \big) \ \ \ \ \ (x>0),
\end{split} 
\end{align}
where $I$ is the modified Bessel function of the first kind. Therefore, we further simplify Eq.(\ref{eq:fmlt}) as follows.\\
For $\xi_1-\xi_2>0$,
\begin{align} 
\begin{split}
f(m,l,t)&=
\frac{\gamma}{m!l!}\sqrt{\pi}\text{e}^{-\frac{\xi_1+\xi_2}{2}t}t^{m+l} \frac{k^{m+l+1}(k^+)^m(k^-)^l}{(\xi_1-\xi_2)^{m+l+1}} \sqrt{(\xi_1-\xi_2)t}\ I_{m+l+\frac{1}{2}}\big(\frac{\xi_1-\xi_2}{2}t\big).
\end{split} 
\end{align}
If $\xi_2-\xi_1>0$ then,
\begin{align} 
\begin{split}
f(m,l,t)&=
\frac{\gamma}{m!l!}\sqrt{\pi}\text{e}^{-\frac{\xi_1+\xi_2}{2}t}t^{m+l} \frac{k^{m+l+1}(k^+)^m(k^-)^l}{(\xi_2-\xi_1)^{m+l+1}} \sqrt{(\xi_2-\xi_1)t}\ I_{m+l+\frac{1}{2}}\big(\frac{\xi_2-\xi_1}{2}t\big).
\end{split} 
\end{align}
Both cases are written as,
\begin{align} 
\begin{split}
\label{fmlt}
f(m,l,t)&=
\frac{\gamma\sqrt{\pi}}{m!l!}\text{e}^{-\frac{\xi_1+\xi_2}{2}t}t^{m+l} \frac{k^{m+l+1}(k^+)^m(k^-)^l}{|\xi_2-\xi_1|^{m+l+1}} \sqrt{|\xi_2-\xi_1|t}\ I_{m+l+\frac{1}{2}}\big(\frac{|\xi_2-\xi_1|}{2}t\big).
\end{split} 
\end{align}
We finally obtain the expression for the velocity distribution. \\
For $m-l>0$,
\begin{align} 
\begin{split}
P(v>0)&=\sum_{\substack{m,l\\m>l}}^{\infty}\int_{0}^{\infty}f(m,l,t)\delta(v-\frac{m-l}{t})dt\\ 
&=\sum_{\substack{m,l\\m>l}}^{\infty}\frac{m-l}{v^2} \frac{\gamma\sqrt{\pi}}{m!l!}\text{e}^{-\frac{\xi_1+\xi_2}{2}\frac{m-l}{v}}\big(\frac{m-l}{v}\big)^{m+l+\frac{1}{2}} \frac{k^{m+l+1}(k^+)^m(k^-)^l}{|\xi_2-\xi_1|^{m+l+\frac{1}{2}}} \ I_{m+l+\frac{1}{2}}\big(\frac{|\xi_2-\xi_1|}{2}\frac{m-l}{v}\big).
\end{split} 
\end{align}
For $m-l<0$,
\begin{align} 
\begin{split}
\label{eq:Pvn}
P(v<0)&=\sum_{\substack{m,l\\l>m}}^{\infty}\int_{0}^{\infty}f(m,l,t)\delta(v-\frac{m-l}{t})dt\\ 
&=\sum_{\substack{m,l\\l>m}}^{\infty}\frac{l-m}{v^2} \frac{\gamma\sqrt{\pi}}{m!l!}\text{e}^{-\frac{\xi_1+\xi_2}{2}\frac{m-l}{v}}\big(\frac{m-l}{v}\big)^{m+l+\frac{1}{2}} \frac{k^{m+l+1}(k^+)^m(k^-)^l}{|\xi_2-\xi_1|^{m+l+\frac{1}{2}}} \ I_{m+l+\frac{1}{2}}\big(\frac{|\xi_2-\xi_1|}{2}\frac{m-l}{v}\big).
\end{split} 
\end{align}
Where $\xi_1=k$ and $\xi_2=k^++k^-+\gamma$. \\

Let us now consider the case $\xi_1=\xi_2$. Let $\xi_1=\xi_2\equiv\xi$, then we need to evaluate the following integral,
\begin{align} 
\begin{split}
\frac{1}{2\pi i}\int_{c-i\infty}^{c+i\infty}\frac{1}{(s+\xi)^{2(n+1)}}\text{e}^{st}ds.
\end{split} 
\end{align}
In this case, we have a single pole of order $2(n+1)$. The result for $f(m,l,t)$ becomes, 
\begin{align} 
\begin{split}
\label{eq:fmlt_degene}
f(m,l,t)=\frac{(m+l)!}{m!l!}k^{m+l+1}(k^+)^m(k^-)^l\gamma\frac{t^{2(m+l)+1}\text{e}^{-\xi t}}{(2(m+l)+1)!},
\end{split} 
\end{align}
leading to
\begin{align} 
\begin{split}
\label{eq:fvp_degene}
P(v>0)=\sum_{\substack{m,l\\m>l}}^{\infty}\frac{m-l}{v^2}\frac{(m+l)!}{m!l!}\frac{k^{m+l+1}(k^+)^m(k^-)^l\gamma}{(2(m+l)+1)!}\Big(\frac{m-l}{v}\Big)^{2(m+l)+1}\text{e}^{-\xi \frac{m-l}{v}},
\end{split} 
\end{align}
\begin{align} 
\begin{split}
\label{eq:fvp_degene}
P(v<0)=\sum_{\substack{m,l\\l>m}}^{\infty}\frac{l-m}{v^2}\frac{(m+l)!}{m!l!}\frac{k^{m+l+1}(k^+)^m(k^-)^l\gamma}{(2(m+l)+1)!}\Big(\frac{m-l}{v}\Big)^{2(m+l)+1}\text{e}^{-\xi \frac{m-l}{v}}.
\end{split} 
\end{align}
Note that Eqs.(\ref{eq:fmlt_degene})-(\ref{eq:fvp_degene}) could also be obtained by taking the limit for $\xi_1 \rightarrow \xi_2$ in Eqs.(\ref{fmlt})-(\ref{eq:Pvn}), knowing that the limit for small argument of the modified Bessel function of the first kind is given in Eq.(10.30.1) of \cite{NISTHandbook}.

For the kinetic scheme in Fig.1(c) and (d) in the main text, the expressions for the velocity distributions are identical. However, the rate $k$, $k^+$, $k^-$, and $\gamma$ depend on ATP concentration in a different manner for the two models, which describe two distinct waiting states of kinesin for ATP.



\section{Randomness parameter}
Randomness parameter, which in some sense is easier to measure in experiments, is a useful way to estimate the number of rate limiting steps of molecular motors~\cite{svoboda1994fluctuation,schnitzer1995statistical}. We here discuss two types of randomness parameters for our models in the main text, chemical randomness $r_C$ and mechanical randomness $r_M$, which are connected by a relation denoted by $\rchem$ as shown below. 
\subsection{Chemical randomness parameter, $\boldsymbol{r_C}$}
Chemical randomness is given  by the following expression,
\begin{align} 
\begin{split}
r_C=\frac{\langle \tau^2 \rangle-\langle \tau \rangle^2} {\langle \tau \rangle^2},
\end{split}                 
\end{align}
where $\tau$ is the dwell time of the motor at a given site and the bracket denotes an average over an ensemble of motors.
Dwell time distributions for stepping forward (\small$+$\normalsize) or backward (\small$-$\normalsize) at a site at time $t$ in our models [Fig.1(c) and (d) in the main text] are given by, 
\small
\begin{align} 
\begin{split}
f_\pm(t)&=\int_{0}^{t}dt^\prime k\text{e}^{-kt^\prime}k^\pm\text{e}^{-(k^++k^-+\gamma)(t-t^\prime)}\\
&=\frac{kk^\pm}{k^++k^-+\gamma-k}(\text{e}^{-kt}-\text{e}^{-(k^++k^-+\gamma)t}). \\ 
\end{split}                 
\end{align}
\normalsize
Thus, the first and second moment of dwell time distribution conditioned by forward or backward step are, 
\small
\begin{align} 
\begin{split}
\langle \tau_\pm \rangle =&\frac{\int_{0}^{\infty}\tau_\pm f_\pm(\tau_\pm)d\tau_\pm}{\int_{0}^{\infty}f_\pm(\tau_\pm)d\tau_\pm}=\frac{k+k^++k^-+\gamma}{k(k^++k^-+\gamma)},
\end{split}                 
\end{align}
\normalsize
\small 
\begin{align} 
\begin{split}
\langle \tau_\pm^2 \rangle =&\frac{\int_{0}^{\infty}\tau_\pm^2f _\pm(\tau_\pm)d\tau_+}{\int_{0}^{\infty}f_\pm(\tau_\pm)d\tau_\pm} \\
=&2\frac{k^2+k(k^++k^-+\gamma)+(k^++k^-+\gamma)^2}{k^2(k^++k^-+\gamma)^2}.
\end{split}                 
\end{align}
\normalsize
Thus, $(\langle \tau_+^2 \rangle-\langle \tau_+ \rangle^2)/ \langle \tau_+ \rangle^2=(\langle \tau_-^2 \rangle-\langle \tau_- \rangle^2)/ \langle \tau_- \rangle^2$, which allows us to express $r_C$ as 
\begin{align} 
\begin{split}
r_C=\frac{k^2+(k^++k^-+\gamma)^2}{(k+k^++k^-+\gamma)^2}.
\end{split}                 
\end{align}
\subsection{Mechanical randomness parameter, $\boldsymbol{r_M}$}
\label{deri_rmec}
We define $r_M$ as,
\begin{align} 
\begin{split}
\label{rmec}
r_M=\lim_{t \to \infty}\frac{\langle n^2(t)\rangle  -\langle  n(t)\rangle^2}{\langle n(t)\rangle},
\end{split} 
\end{align}
where $n(t)$ is the position of the motor at time $t$. 
In our model, we can obtain the expression for the probability distribution that the motor is at site $n$ at time $t$, which is needed to calculate the moments in (\ref{rmec}).\\
Using (\ref{fmlt}) we obtain the probability that the motor takes $n = (m-l)$ , 
\begin{align} 
\begin{split}
\label{fnt}
f(n,t)=&\sum_{m,l}\delta_{m,n+l}f(m,l,t)\\
=&\sum_{l=0}^{\infty}\frac{\gamma\sqrt{\pi}}{(n+2l)!l!}\text{e}^{-\frac{\xi_1+\xi_2}{2}t}t^{n+2l+\frac{1}{2}} \frac{k^{n+2l+1}(k^+)^{n+l}(k^-)^l}{|\xi_2-\xi_1|^{n+2l+\frac{1}{2}}} \ I_{n+2l+\frac{1}{2}}\big(\frac{|\xi_2-\xi_1|}{2}t\big).
\end{split}                 
\end{align}


However, as written the sum (\ref{fnt}) accounts for contributions from motors that have detached from the track before sufficiently long time $t$ has elapsed. The appropriate probability distribution to get the moments in Eq.(\ref{rmec}) is the re-normalized probability distribution at each time $t$, which accounts only for the motors which stay on the track for a long time $t$. We denote this probability distribution as $\bar{f}$, which is defined as,\\
\begin{align} 
\begin{split}
\label{eq:fbarp}
\bar{f}(n,t)
=&\frac{1}{C}\sum_{l=0}^{\infty}\frac{\gamma\sqrt{\pi}}{(n+2l)!l!}\text{e}^{-\frac{\xi_1+\xi_2}{2}t}t^{n+2l+\frac{1}{2}} \frac{k^{n+2l+1}(k^+)^{n+l}(k^-)^l}{|\xi_2-\xi_1|^{n+2l+\frac{1}{2}}} \ I_{n+2l+\frac{1}{2}}\big(\frac{|\xi_2-\xi_1|}{2}t\big),
\end{split}                 
\end{align}

where 
\begin{align} 
\begin{split}
\label{eq:Const}
C
=&\sum_{l=0}^{\infty}\sum_{n=-\infty}^{\infty}\frac{\gamma\sqrt{\pi}}{(n+2l)!l!}\text{e}^{-\frac{\xi_1+\xi_2}{2}t}t^{n+2l+\frac{1}{2}} \frac{k^{n+2l+1}(k^+)^{n+l}(k^-)^l}{|\xi_2-\xi_1|^{n+2l+\frac{1}{2}}}  I_{n+2l+\frac{1}{2}}\big(\frac{|\xi_2-\xi_1|}{2}t\big).
\end{split}                 
\end{align}
 We used the distribution Eq.(\ref{eq:fbarp}) to obtain the first and second moment of $n$ for the calculation of mechanical randomness parameter.
In practice, we calculated the first and second moments $\langle  n(t)\rangle$ and $\langle n^2(t)\rangle$ at $t=0.5$s, which is long enough for  $\langle  n(t)\rangle$ to decay significantly. The double summation in Eq.(\ref{eq:Const}) should include enough terms to ensure  convergence of  $\langle  n(t)\rangle$ and $\langle n^2(t)\rangle$. We truncated the summation at 30 and 130 for $l$ and $n$, respectively. Needless to say that if $t$ is extended beyond $0.5$s then a larger number of terms will have to be calculated to obtain converged results.
\begin{figure}[]
\begin{center}
\includegraphics[width=\textwidth]{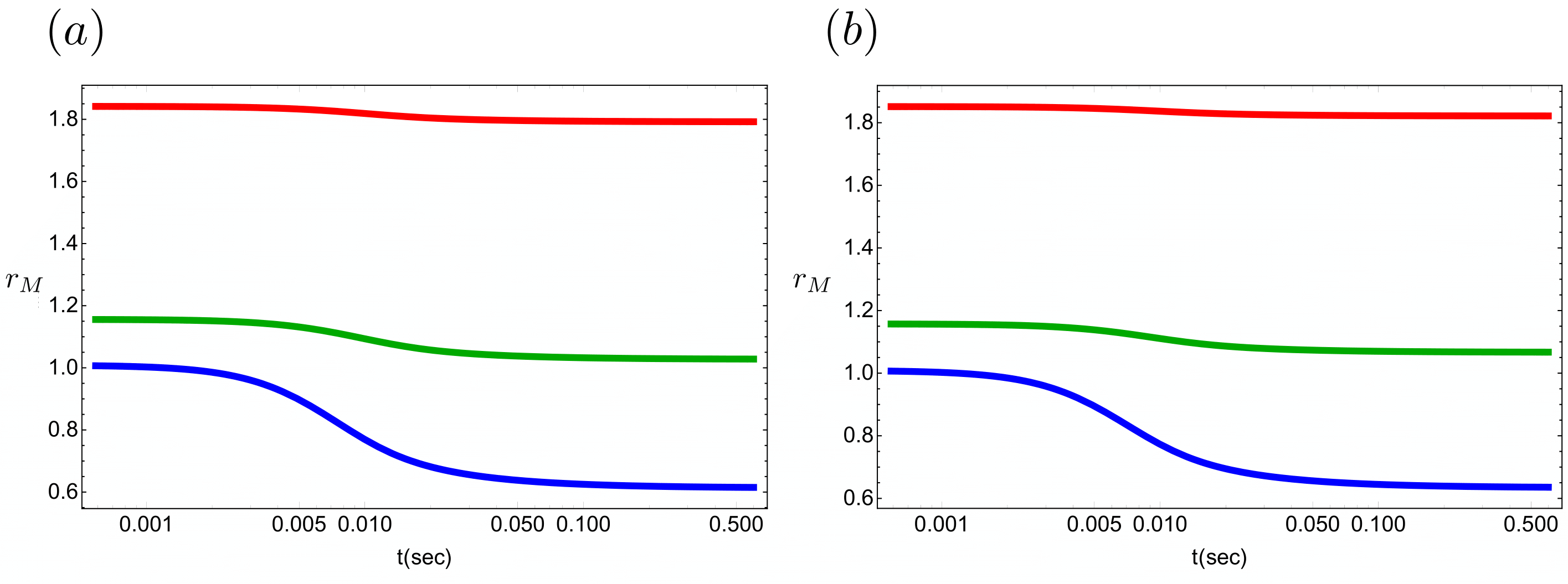}
\end{center}
\caption{\label{fig:randomness_relax}Relaxation of mechanical randomness parameter at $F=0\pN$ (blue), $F=4\pN$ (green), and $F=6\pN$ (red). [T]=1mM in all cases. (a) 2HB model. (b) 1HB model. In all cases $t=0.5$s is long enough for $r_M$ to have a plateau value.}
\end{figure}

If we denote $\rchem$ as the chemical randomness parameter, which takes backward steps into account, we found that,

\begin{align} 
\begin{split}
\label{eq:rchem0}
\rchem=\frac{(2P_+-1)r_M-4P_+(1-P_+)}{(2P_+-1)^2}.
\end{split}                 
\end{align}
Thus, $\rchem$ can be calculated from mechanical randomness parameter by taking into account backward steps. The derivation of this equation can be found in the literature in a different context~\cite{shaevitz2005statistical,chemla2008exact}. We show that in the next section this relation can be derived by including backward steps in the work by Schnitzer and Block~\cite{schnitzer1995statistical}.
In order to calculate $\rchem$ we need to compute the probability of forward step, $P_+$. We assume $P_++P_-=1$ for all times until the motor detaches.\\
Thus, in our model,
\begin{align} 
\begin{split}
P_+=\frac{\int_{0}^{\infty}f_+(\tau_+)d\tau_+}{\int_{0}^{\infty}f_+(\tau_+)d\tau_++\int_{0}^{\infty}f_-(\tau_-)d\tau_-}=\frac{k^+}{k^++k^-}.
\end{split}                 
\end{align}
Using $r_M$ and $P_+$, we are able to calculate $\rchem$.

\subsection{Derivation of mechanical randomness parameter with backward steps}
We derive Eq.(\ref{eq:rchem0}) by incorporating backward steps into the work by Schnitzer and Block~\cite{schnitzer1995statistical}. We denote the dwell time distribution corresponding to forward and backward steps as $f_+(t)$ and $f_-(t)$, respectively. Let $g(m,l,t)$ be the probability distribution of taking $m$ forward steps and $l$ backward steps before time $t$.
The Laplace transform of $g(m,l,t)$, is written as follows,
\begin{align} 
\begin{split}
\label{eq:fmls}
\tilde{g}(m,l,s) =& \frac{(m+l)!}{m!l!}(\tilde{f}_+(s))^{m}(\tilde{f}_-(s))^{l}\frac{1-\tilde{f}_+(s)-\tilde{f}_-(s)}{s}.\\ 
\end{split} 
\end{align}
The last term in (\ref{eq:fmls}) is the Laplace transform of $1-\int_{0}^{t}(f_+(t)+f_-(t))dt$, means neither forward nor backward step occurs in the last time step.
For $n\equiv m-l\geq0$,
\begin{align} 
\begin{split}
\tilde{g}_+(n,s) =&\sum_{m,l}\delta_{n,m-l}\tilde{g}(m,l,s) \\
=&[\tilde{f}_+(s)]^n \ _2\text{F}_1(\frac{1+n}{2},\frac{2+n}{2},1+n;4\tilde{f}_+(s)\tilde{f}_-(s)) \frac{1-\tilde{f}_+(s)-\tilde{f}_-(s)}{s}.
\end{split} 
\end{align}
Using the special case of Hypergeometric function (page 556 of \cite{Handbook_math}) we obtain,
\begin{align} 
\begin{split}
\tilde{g}_+(n,s) =\Biggl[\frac{2\tilde{f}_+(s)}{1+\sqrt{1-4\tilde{f}_+(s)\tilde{f}_-(s)}}\Biggl]^n\frac{1}{\sqrt{1-4\tilde{f}_+(s)\tilde{f}_-(s)}}\frac{1-\tilde{f}_+(s)-\tilde{f}_-(s)}{s}.
\end{split} 
\end{align}
For  $n\leq0$, a similar procedure leads to, 
\begin{align} 
\begin{split}
\tilde{g}_-(n,s) =\Biggl[\frac{2\tilde{f}_-(s)}{1+\sqrt{1-4\tilde{f}_+(s)\tilde{f}_-(s)}}\Biggl]^{-n}\frac{1}{\sqrt{1-4\tilde{f}_+(s)\tilde{f}_-(s)}}\frac{1-\tilde{f}_+(s)-\tilde{f}_-(s)}{s}.
\end{split} 
\end{align}
Thus, 
\begin{align} 
\begin{split}
\tilde{g}_\pm(n,s) =\biggl(\frac{2\tilde{f}_\pm(s)}{1+\sqrt{1-4\tilde{f}_+(s)\tilde{f}_-(s)}}\biggl)^{|n|}\frac{1}{\sqrt{1-4\tilde{f}_+(s)\tilde{f}_-(s)}}\frac{1-\tilde{f}_+(s)-\tilde{f}_-(s)}{s}.
\end{split} 
\end{align}
The first and second moment of $n$ are defined as,
\begin{align} 
\begin{split}
\label{eq:moments}
\langle n(s)\rangle=&\sum_{n=0}^{\infty}n\tilde{g}_+(n,s)+\sum_{n=-\infty}^{0}n\tilde{g}_-(n,s) \\
\equiv& \langle n(s)\rangle_+ +\langle n(s)\rangle_- \\
\langle n^2(s)\rangle=&\sum_{n=0}^{\infty}n^2\tilde{g}_+(n,s)+\sum_{n=-\infty}^{0}n^2\tilde{g}_-(n,s) \\
\equiv& \langle n^2(s)\rangle_+ +\langle n^2(s)\rangle_- .
\end{split} 
\end{align}
It is convenient to use the generating function of $\tilde{g}_\pm(n,s)$ to calculate the moments, and it is a simple geometric sum given by,
\begin{align} 
\begin{split}
Z_{+}(x,s)=&\sum_{n=0}^{\infty}\tilde{g}_+(n,s)x^n\\
=&\frac{1}{1-\frac{2\tilde{f}_+(s)x}{1+\sqrt{1-4\tilde{f}_+(s)\tilde{f}_-(s)}}}\frac{1}{\sqrt{1-4\tilde{f}_+(s)\tilde{f}_-(s)}}\frac{1-\tilde{f}_+(s)-\tilde{f}_-(s)}{s}.
\end{split} 
\end{align}
The average number of forward step is calculated using the generating function as,
\begin{align} 
\begin{split}
\label{avenp}
\langle n(s)\rangle_+=&\frac{\partial Z_+}{\partial x}\Big|_{x=1}\\ 
=&\frac{2\tilde{f}_+(s)\Big[1+\sqrt{1-4\tilde{f}_+(s)\tilde{f}_-(s)}\ \Big]}{\sqrt{1-4\tilde{f}_+(s)\tilde{f}_-(s)}\biggl[1-2\tilde{f}_+(s)+\sqrt{1-4\tilde{f}_+(s)\tilde{f}_-(s)}\ \biggl]^2}\frac{1-\tilde{f}_+(s)-\tilde{f}_-(s)}{s}.
\end{split} 
\end{align}
The second derivative of the generating function leads to,
\begin{align} 
\begin{split}
\langle n^2(s)\rangle_+-\langle n(s)\rangle_+=&\frac{\partial^2 Z_+}{\partial x^2}\Big|_{x=1}\\ 
=&\frac{8[\tilde{f}_+(s)]^2\Big[1+\sqrt{1-4\tilde{f}_+(s)\tilde{f}_-(s)}\ \Big]}{\sqrt{1-4\tilde{f}_+(s)\tilde{f}_-(s)}\biggl[1-2\tilde{f}_+(s)+\sqrt{1-4\tilde{f}_+(s)\tilde{f}_-(s)}\ \biggl]^3}\frac{1-\tilde{f}_+(s)-\tilde{f}_-(s)}{s}.
\end{split} 
\end{align}
Thus,
\begin{align} 
\begin{split}
\label{aven2p}
\langle n^2(s)\rangle_+
=&\frac{2\tilde{f}_+\Big[1+\sqrt{1-4\tilde{f}_+(s)\tilde{f}_-(s)}\ \Big]\Big[1+2\tilde{f}_+(s)+\sqrt{1-4\tilde{f}_+(s)\tilde{f}_-(s)}\ \Big]}{\sqrt{1-4\tilde{f}_+(s)\tilde{f}_-(s)}\biggl[1-2\tilde{f}_+(s)+\sqrt{1-4\tilde{f}_+(s)\tilde{f}_-(s)}\ \biggl]^3}\frac{1-\tilde{f}_+(s)-\tilde{f}_-(s)}{s}.
\end{split} 
\end{align}
In a similar manner,
\begin{align} 
\begin{split}
\label{avenn}
\langle n(s)\rangle_- =-\frac{2\tilde{f}_-(s)\Big[1+\sqrt{1-4\tilde{f}_+(s)\tilde{f}_-(s)}\ \Big]}{\sqrt{1-4\tilde{f}_+(s)\tilde{f}_-(s)}\biggl[1-2\tilde{f}_-(s)+\sqrt{1-4\tilde{f}_+(s)\tilde{f}_-(s)}\biggl]^2}\frac{1-\tilde{f}_+(s)-\tilde{f}_-(s)}{s},
\end{split} 
\end{align}
\begin{align} 
\begin{split}
\label{aven2n}
\langle n^2(s)\rangle_-
=&\frac{2\tilde{f}_-\Big[1+\sqrt{1-4\tilde{f}_+(s)\tilde{f}_-(s)}\ \Big]\Big[1+2\tilde{f}_+(s)+\sqrt{1-4\tilde{f}_+(s)\tilde{f}_-(s)}\ \Big]}{\sqrt{1-4\tilde{f}_+(s)\tilde{f}_-(s)}\biggl[1-2\tilde{f}_-(s)+\sqrt{1-4\tilde{f}_+(s)\tilde{f}_-(s)}\ \biggl]^3}\frac{1-\tilde{f}_+(s)-\tilde{f}_-(s)}{s}.
\end{split} 
\end{align}
By substituting the above expressions in Eq.(\ref{eq:moments}) we obtain,
\small
\begin{align} 
\begin{split}
\label{aven}
\langle n(s)\rangle
=\frac{2\Big[1+\sqrt{1-4\tilde{f}_+(s)\tilde{f}_-(s)}\ \Big]}{\sqrt{1-4\tilde{f}_+(s)\tilde{f}_-(s)}} \frac{1-\tilde{f}_+(s)-\tilde{f}_-(s)}{s}  &\Biggl[\frac{\tilde{f}_+(s)}{\Big[1-2\tilde{f}_+(s)+\sqrt{1-4\tilde{f}_+(s)\tilde{f}_-(s)}\ \Big]^2} \\ &- \frac{\tilde{f}_-(s)}{\Big[1-2\tilde{f}_-(s)+\sqrt{1-4\tilde{f}_+(s)\tilde{f}_-(s)}\ \Big]^2} \Biggl], 
\end{split} 
\end{align}
\normalsize
\small
\begin{align} 
\begin{split}
\label{aven2}
\langle n^2(s)\rangle
=\frac{2\Big[1+\sqrt{1-4\tilde{f}_+(s)\tilde{f}_-(s)}\ \Big]}{\sqrt{1-4\tilde{f}_+(s)\tilde{f}_-(s)}} \frac{1-\tilde{f}_+(s)-\tilde{f}_-(s)}{s}  &\Biggl[\frac{\tilde{f}_+(s)\Big[1+2\tilde{f}_+(s)+\sqrt{1-4\tilde{f}_+(s)\tilde{f}_-(s)}\ \Big]}{\Big[1-2\tilde{f}_+(s)+\sqrt{1-4\tilde{f}_+(s)\tilde{f}_-(s)}\ \Big]^3}  \\ &+ \frac{\tilde{f}_-(s)\Big[1+2\tilde{f}_-(s)+\sqrt{1-4\tilde{f}_+(s)\tilde{f}_-(s)}\ \Big]}{\Big[1-2\tilde{f}_-(s)+\sqrt{1-4\tilde{f}_+(s)\tilde{f}_-(s)}\ \Big]^3} \Biggl] .
\end{split} 
\end{align}
\normalsize
We express $\tilde{f}_\pm(s)$ using Taylor expansion,
\begin{align} 
\begin{split}
\label{taylor}
\tilde{f}_+(s)=&P_+\sum_{k=0}^{\infty}\frac{\langle \tau_+^k \rangle(-s)^k}{k!}\\ 
\tilde{f}_-(s)=&P_-\sum_{k=0}^{\infty}\frac{\langle \tau_-^k \rangle(-s)^k}{k!}.  
\end{split} 
\end{align}
The moments of $\tilde{f}_\pm(t)$ are given by,
\begin{align} 
\begin{split}
\label{moments}
\langle \tau_+^k \rangle =&\frac{\int_{0}^{\infty}\tau_+^kf_+(\tau_+)d\tau_+}{\int_{0}^{\infty}f_+(\tau_+)d\tau_+}=\frac{\int_{0}^{\infty}\tau_+^kf_+(\tau_+)d\tau_+}{P_+} =\frac{(-1)^k\frac{d^k\tilde{f}_+(s)}{ds^k}}{P_+}, \\ 
\langle \tau_-^k \rangle =&\frac{\int_{0}^{\infty}\tau_-^kf_-(\tau_-)d\tau_-}{\int_{0}^{\infty}f_-(\tau_-)d\tau_-}=\frac{\int_{0}^{\infty}\tau_-^kf_-(\tau_-)d\tau_-}{P_-} =\frac{(-1)^k\frac{d^k\tilde{f}_-(s)}{ds^k}}{P_-}.
\end{split} 
\end{align}
From here we assume the relation $P_+ + P_-=1$.
Substituting (\ref{taylor}) into (\ref{aven}) and (\ref{aven2}) with the condition either $\frac{1}{2} < P_+ \leq 1$ or  $0 \leq P_+ < \frac{1}{2}$ both yield,
\begin{align} 
\begin{split}
\langle n(s)\rangle=&\frac{2P_+-1}{\big(P_+\langle \tau_+ \rangle +(1-P_+)\langle \tau_- \rangle  \big)s^2} \\ 
+&\frac{-2P_+^2\langle \tau_+ \rangle^2 + 2(1-P_+)^2\langle \tau_- \rangle^2 + P_+(2P_+ - 1)\langle \tau_+^2 \rangle + (2P_+-1)(1-P_+)\langle \tau_-^2 \rangle }  {2\big(P_+\langle \tau_+ \rangle +(1-P_+)\langle \tau_- \rangle  \big)^2s} 
+ O(1),
\end{split} 
\end{align}
and 
\footnotesize
\begin{align} 
\begin{split}
\langle n^2(s)\rangle=&\frac{2(2P_+-1)^2}{\big(P_+\langle \tau_+ \rangle +(1-P_+)\langle \tau_- \rangle  \big)^2s^3} +\\ 
&\frac{-P_+^2(8P_+-5)\langle \tau_+ \rangle^2 + (1-P_+)^2(8P_+-3)\langle \tau_- \rangle^2 +  2(2P_+-1)^2 \big((1-P_+)\langle \tau_-^2 \rangle +P_+\langle \tau_+^2 \rangle\big)  + 2(1-P_+)P_+\langle \tau_+ \rangle \langle \tau_- \rangle}  {\big(P_+\langle \tau_+ \rangle +(1-P_+)\langle \tau_- \rangle  \big)^3s^2} 
\\& + O(s^{-1}).
\end{split} 
\end{align}
\normalsize
After inverse Laplace transform,
\begin{align} 
\begin{split}
\langle n(t)\rangle=&\frac{2P_+-1}{\big(P_+\langle \tau_+ \rangle +(1-P_+)\langle \tau_- \rangle  \big)}t \\ 
+&\frac{-2P_+^2\langle \tau_+ \rangle^2 + 2(1-P_+)^2\langle \tau_- \rangle^2 + P_+(2P_+ - 1)\langle \tau_+^2 \rangle + (2P_+-1)(1-P_+)\langle \tau_-^2 \rangle }  {2\big(P_+\langle \tau_+ \rangle +(1-P_+)\langle \tau_- \rangle  \big)^2} 
+ O(t^{-1}), \\ 
\end{split} 
\end{align}
and
\begin{align} 
\begin{split}
\langle n^2(t)\rangle  -\langle  n(t)\rangle^2=&\frac{P_+(2P_+-1)^2\langle \tau_+^2 \rangle-P_+^2(4P_+-3)\langle \tau_+ \rangle^2}{\big(P_+\langle \tau_+ \rangle +(1-P_+)\langle \tau_- \rangle  \big)^3}t \\ 
& +\frac{(1-P_+)(2P_+-1)^2\langle \tau_-^2 \rangle-(1-P_+)^2(1-4P_+)\langle \tau_- \rangle^2}{\big(P_+\langle \tau_+ \rangle +(1-P_+)\langle \tau_- \rangle  \big)^3}t \\ 
& + \frac{2(1-P_+)P_+\langle \tau_+ \rangle \langle \tau_- \rangle}{\big(P_+\langle \tau_+ \rangle +(1-P_+)\langle \tau_- \rangle  \big)^3}t 
+ O(1).
\end{split} 
\end{align}
Now we compute the randomness parameter as 
\begin{align} 
\begin{split}
r_M=&\lim_{t \to \infty}\frac{\langle n^2(t)\rangle  -\langle  n(t)\rangle^2}{\langle n(t)\rangle} \\ 
=&\ \ \frac{P_+(2P_+-1)^2\langle \tau_+^2 \rangle-P_+^2(4P_+-3)\langle \tau_+ \rangle^2}{\big(P_+\langle \tau_+ \rangle +(1-P_+)\langle \tau_- \rangle  \big)^2(2P_+-1)} \\ 
& +\frac{(1-P_+)(2P_+-1)^2\langle \tau_-^2 \rangle-(1-P_+)^2(1-4P_+)\langle \tau_- \rangle^2}{\big(P_+\langle \tau_+ \rangle +(1-P_+)\langle \tau_- \rangle  \big)^2(2P_+-1)} \\ 
& + \frac{2(1-P_+)P_+\langle \tau_+ \rangle \langle \tau_- \rangle}{\big(P_+\langle \tau_+ \rangle +(1-P_+)\langle \tau_- \rangle  \big)^2(2P_+-1)} .
\end{split} 
\end{align}
For the special case, namely $\langle \tau_+ \rangle=\langle \tau_- \rangle$ and $\langle \tau^2_+ \rangle=\langle \tau^2_- \rangle$, $r_M$ in the above equation reduces to the following expression,
\begin{align} 
\begin{split}
\label{rsimp}
r_M=\frac{(2P_+-1)^2\langle \tau^2 \rangle-(8P_+^2-8P_++1)\langle \tau \rangle^2}{(-1+2P_+)\langle \tau \rangle^2}.
\end{split} 
\end{align}
Manipulation of Eq.($\ref{rsimp}$) leads to, 
\begin{align} 
\begin{split}
r_C=\frac{(2P_+-1)r_M-4P_+(1-P_+)}{(2P_+-1)^2}\equiv \rchem,
\end{split} 
\end{align}
where $r_C=\frac{\langle \tau^2 \rangle-\langle \tau \rangle^2}{\langle \tau \rangle^2}$. Note that $r_C=r_M$ if $P_+$ is unity, which holds in the  absence of backward steps. 

\section{Two models for how kinesin waits for ATP}
\begin{figure}[H]
\begin{center}
\includegraphics[width=\textwidth]{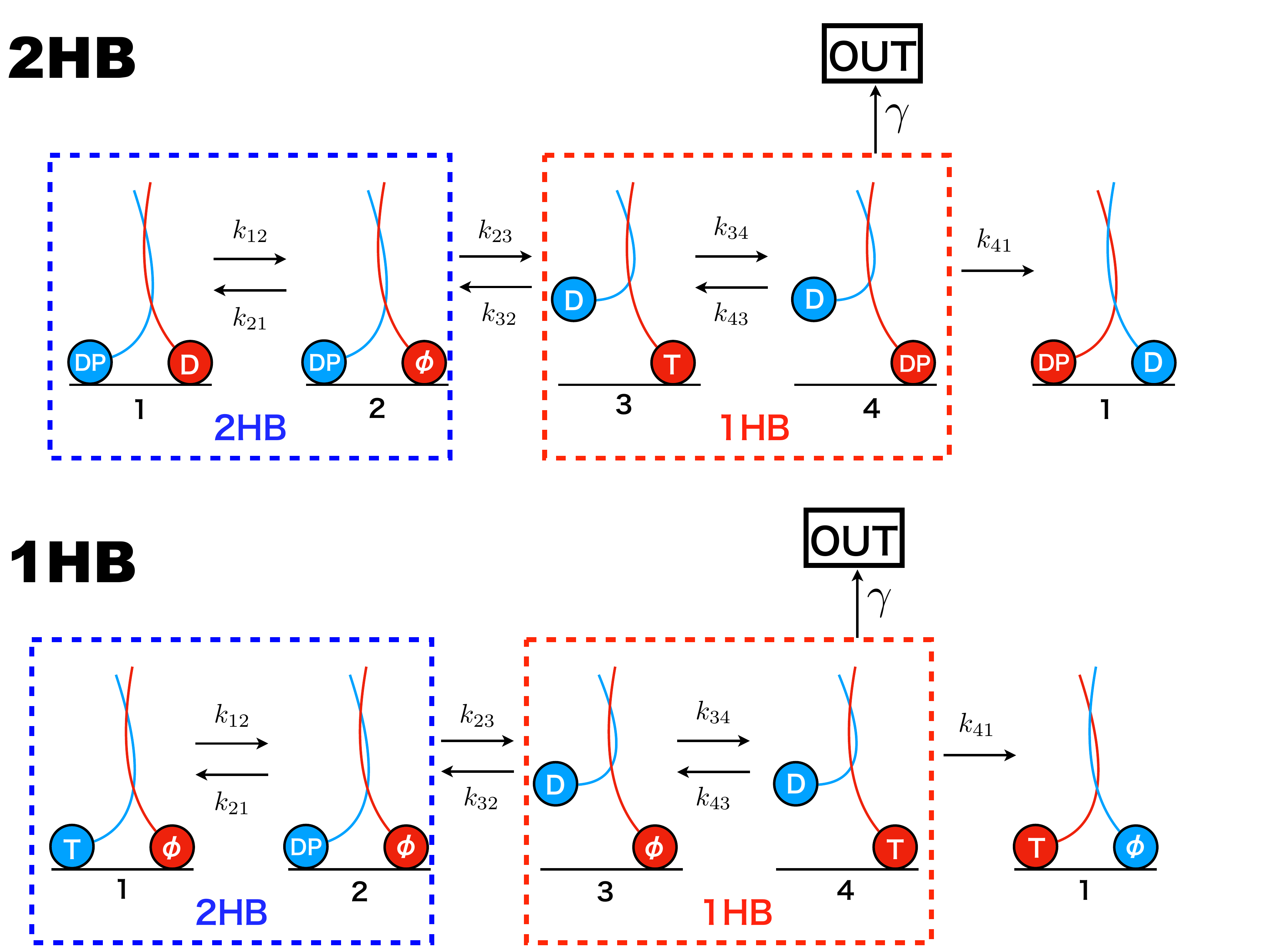}
\end{center}
\caption{\label{Fig:Map} A sketch showing all the relevant rates involved in the stepping of kinesin. The upper (lower) panel shows the  2HB (1HB) waiting state. In the 2HB model , ATP binds to the LH when both heads are bound to the microtuble (MT). In the 1HB model ATP binding occurs only after the trailing head detaches from the MT. At extremely low ATP concentration the 1HB model is more likely.}
\end{figure}
\begin{figure}[]
\begin{center}
\includegraphics[width=0.5\textwidth]{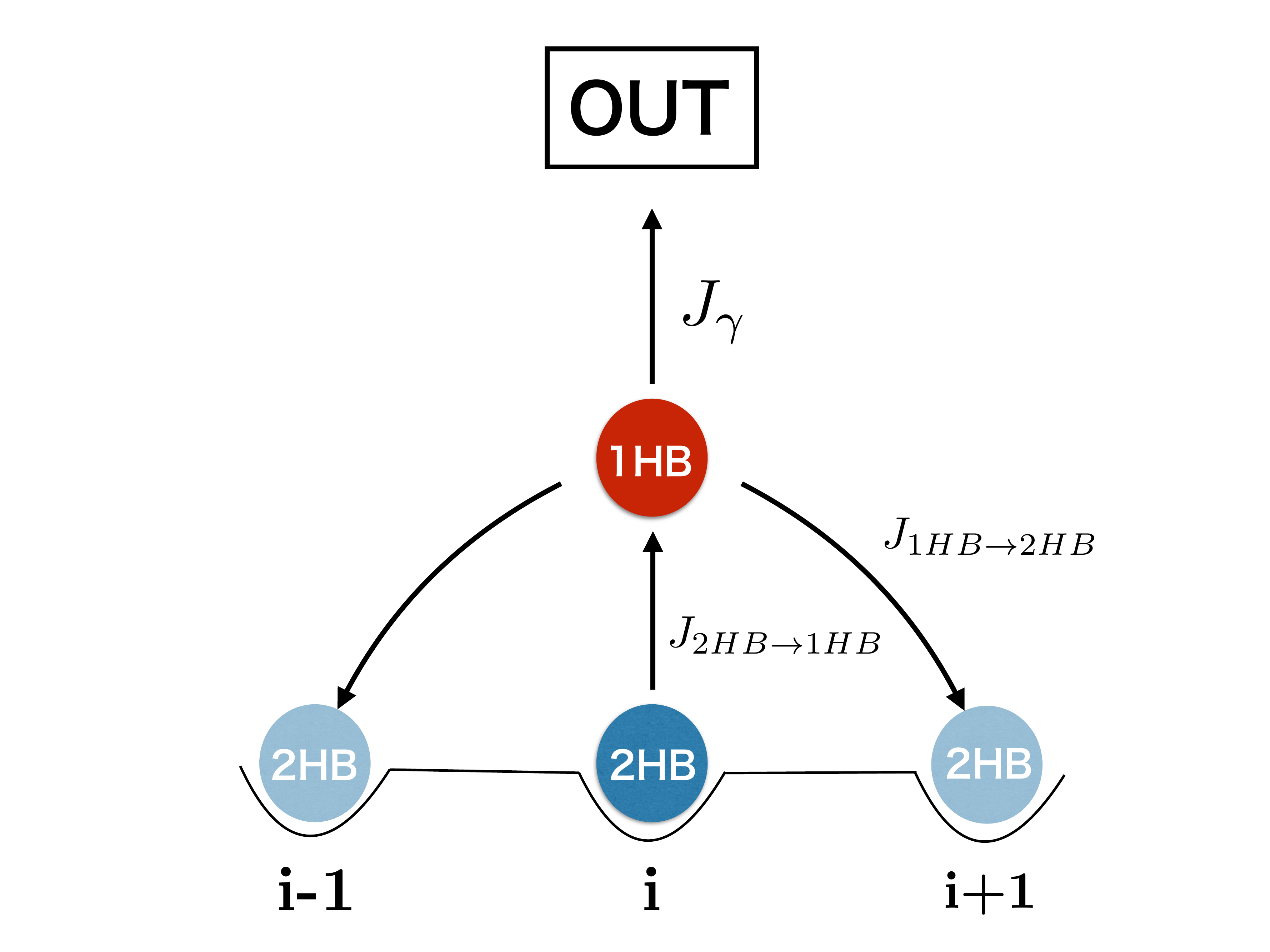}
\end{center}
\caption{\label{Fig:2stateswoBS} Simplified two states model.}
\end{figure}
As explained in the main text, the two models (Fig.\ref{Fig:2stateswoBS}) have been proposed for the waiting states of kinesin for ATP binding. In the 2HB model, ATP binds to the leading head when both the heads are bound to the MT (upper panel in Fig.\ref{Fig:Map}). In contrast, Isojima {\it et al.} have suggested, based on dark field microscopy, that ATP binds only after the trailing head detaches leading to the so-called vulnerable state (Fig.\ref{Fig:2stateswoBS} bottom panel).  
In order to simplify the kinetic scheme, we merge the 4 states in Fig.\ref{Fig:Map} into two states shown in Fig.\ref{Fig:2stateswoBS} in order to calculate the net flux of transition from 2HB state to 1HB state ($J_{2HB \rightarrow 1HB}$), forward step ($J_{1HB \rightarrow 2HB}$), and detachment ($J_{\gamma}$). The simplification allows us to obtain closed form expressions for $J_{2HB \rightarrow 1HB}$, $J_{1HB \rightarrow 2HB},$ and $J_{\gamma}$. 

In the following calculations, we employ method pioneered by Hill~\cite{hillfree,Hill2879}. We may design state 3 to be the absorbing state for the transition from 2HB to 1HB (state 2 to state 3), state 1 to be the absorbing state for the transition from 1HB to 2HB (state 4 to state 1). In addition, detachment is also an absorbing state when kinesin is in the 1HB state. By obtaining the stationary solution of the following sets of master equations, with the conditions $P_1+P_2=1$ and $P_3+P_4=1$,
\begin{align} 
\begin{split}
&\frac{dP_1}{dt}=-k_{12}P_1 + (k_{23}+k_{21})P_2, \\
&\frac{dP_2}{dt}=k_{12}P_1-(k_{21}+k_{23})P_2, \\ 
\end{split} 
\end{align}

\begin{align} 
\begin{split}
&\frac{dP_3}{dt}=-k_{34}P_3 + (k_{43}+k_{41}+\gamma)P_2, \\
&\frac{dP_4}{dt}=k_{34}P_3-(k_{41}+k_{43}+\gamma)P_2,\\ 
\end{split} 
\end{align}
we obtain,
\begin{align} 
\begin{split}
J_{2HB \rightarrow 1HB}&=\frac{k_{12} k_{23}}{k_{12}+k_{21}+k_{23}},\\
J_{1HB \rightarrow 2HB}&=\frac{k_{34} k_{41}}{\gamma +k_{34}+k_{41}+k_{43}},\\
J_{\gamma}&=\frac{\gamma  k_{34}}{\gamma +k_{34}+k_{41}+k_{43}},
\end{split} 
\end{align}
ATP binds with the rate $k_{23}$ in the 2HB waiting state (upper panel in Fig.\ref{Fig:Map}), which we parametrize as $k_{23}=[\text{T}]k^0_{23}$, where [T] is the ATP concentration. On the other hand, ATP binding occurs with rate $k_{34}$ in the 1HB model with detached TH. Thus, we include ATP dependence in $k_{34}=[\text{T}]k^0_{34}$.
With these assumptions, the expressions for the fluxes are given by,
\begin{align} 
\begin{split}
\label{eq:flux2HB}
J_{2HB \rightarrow 1HB}&=\frac{k_{12} [\text{T}]k^0_{23}}{k_{12}+k_{21}+[\text{T}]k^0_{23}},\\
J_{1HB \rightarrow 2HB}&=\frac{k_{34} k_{41}}{\gamma +k_{34}+k_{41}+k_{43}},\\
J_{\gamma}&=\frac{\gamma  k_{34}}{\gamma +k_{34}+k_{41}+k_{43}},
\end{split} 
\end{align}
in the 2HB mdoel. The analogous expressions in the 1HB model are,
\begin{align} 
\begin{split}
\label{eq:flux1HB}
J_{2HB \rightarrow 1HB}&=\frac{k_{12} k_{23}}{k_{12}+k_{21}+k_{23}},\\
J_{1HB \rightarrow 2HB}&=\frac{[\text{T}]k^0_{34} k_{41}}{\gamma +[\text{T}]k^0_{34}+k_{41}+k_{43}},\\
J_{\gamma}&=\frac{\gamma  [\text{T}]k^0_{34}}{\gamma +[\text{T}]k^0_{34}+k_{41}+k_{43}},
\end{split} 
\end{align}
Thus, the Michaelis-menten (MM) kinetics naturally arises from this coarse grained procedure in $J_{2HB \rightarrow 1HB}$ (Eq.\ref{eq:flux2HB}) and $J_{1HB \rightarrow 2HB}$ (Eq.\ref{eq:flux1HB}). Since the MM constant in $J_{1HB \rightarrow 2HB}$ and $J_{\gamma}$ for the 1HB waiting model are identical, we used the functional form 
$k^+= \frac{k_0^+[\text{T}]}{K_T+[\text{T}]}\text{e}^{-\beta F d^+}$, $k^-= \frac{k_0^-[\text{T}]}{K_T+[\text{T}]}\text{e}^{\beta F d^-}$, and $\gamma= \frac{\gamma_0[\text{T}]}{K_T+[\text{T}]}\text{e}^{F/F_d}$. Using identical MM constant in $k^+$, $k^-$, and $\gamma$ in 1HB waiting model leads to ATP independent ratio of probability of forward step, backward step, and detachment. This is realized in 2the HB waiting model as well, and is in accord with experimental observation \cite{nishiyama2002chemomechanical,carter2005mechanics}.   

\section{Variant of 1HB model\Large : \normalsize $k^-$ is independent of [T]}
\begin{figure}[]
\begin{center}
\includegraphics[width=\textwidth]{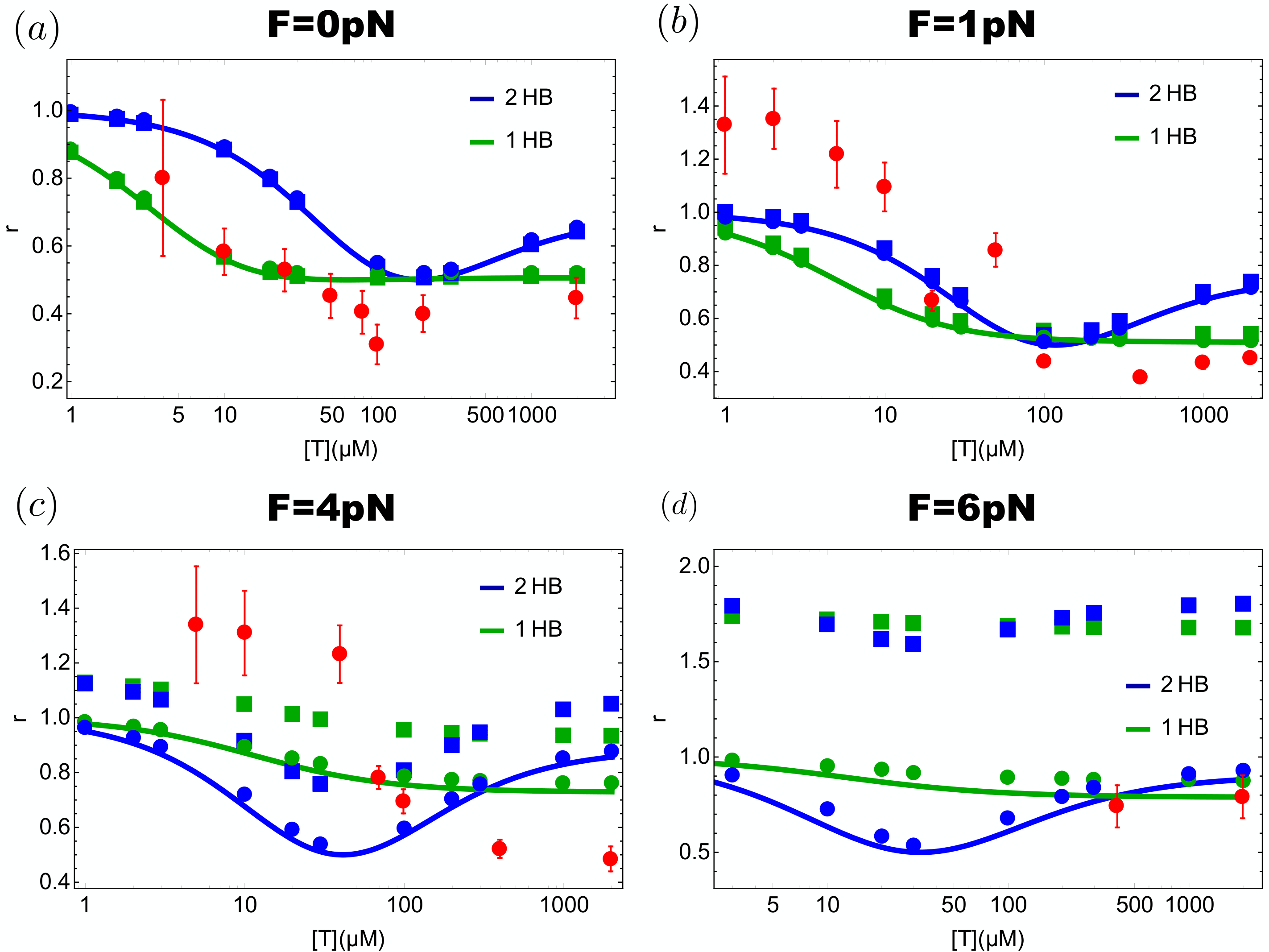}
\end{center}
\caption{\label{fig:randomness_variant}  Theoretical predictions for the ATP concentration dependence of the three randomness parameters, $r_{M}$, $r_{C}$, and $\rchem$ at different external loads for the 2HB model (blue) and the variant of 1HB model (green). Filled circles, filled squares and lines denote $r_M$, $\rchem$, and $r_C$, respectively. Red circles with error bar in (a) are the experimentally measured randomness parameter at $F=0$ in~\cite{verbrugge2009novel}. Red circles with error bar in (b)-(d) are the randomness parameters measured in \cite{visscher1999single}; (b) for 1.05 pN, (c) for 3.59 pN, and (d) for 5.69 pN. As explained in the discussion section in the main text, randomness parameters in our schemes are always equal or grater than 0.5. The results for the 1HB model are obtained by assuming that the rate for the backward step is a constant independent of the ATP concentration. }
\end{figure}

It appears logical that the backward step should be the reverse of the forward step, which implies that it too should occur by a hand-over-hand mechanism with the rate being dependent on ATP. For reasons discussed in the main text, depending on the pathway that kinesin takes to go take a  backward step,  it is possible that rate of backward step $k^-$ does not depend on ATP binding. Therefore,  we created a variant of the 1HB model [Fig.1(c) in the main text] in which $k^-$ is independent of [T]. The load dependence is identical to the original 1HB model, namely $k^-=k^-_0\text{e}^{\beta F^-_d}$. We followed the same procedure described in the main text to obtain the parameters for the variant of 1HB model, which are listed in Table.\ref{Table:variant_1HB}. Interestingly, the values of the many parameters extracted using the 1HB and variant 1HB models are not that dissimilar (compare Table.\ref{Table:variant_1HB} and Table 2 in the main text). We show in Fig.\ref{fig:randomness_variant} that our main prediction about the qualitative behavior of randomness parameters are robust: the randomness parameters for the variant of 1HB waiting model also show monotonic decrease as  [T] is increased.   Thus, regardless of the mechanism of the backward step we conclude that measurements of the randomness parameter as a function of load and ATP concentration using currently available high temporal resolution experiments should resolve the nature of waiting states for ATP.

\begin{table}[t]
\begin{center}
  \begin{tabular}{lllll}
    $k_0$      &$244.0(s^{-1})$  &\ \ \ &$d^+$&$2.2~(\text{nm})$\\
    $k_0^+$    &$303.1(s^{-1})$     &\ \ \            &$d^-$&$0.7(\text{nm})$ \\
    $k_0^-$    &$1.3(s^{-1})$&\ \ \                   &$F_d$&$3.0(\text{pN})$\\
    $\gamma_0$ &$2.4(s^{-1})$&\ \ \                  & $K_T$ &$16.0(\mu \text{M})$ \\
  \end{tabular}
\end{center} 
\caption{\label{Table:variant_1HB}Extracted parameters for the variant of 1HB model}
\end{table}

\newpage
\section{Fitting theory to experimental data}
In order to obtain the parameters for the model, we analyzed the distribution of run length and velocity at $F=0$ using the data reported in Ref.\cite{Walter_2012}. For Kin1, the measured mean velocity at 0 load is 1089 nm/s, which implies $J^+-J^-=132.8$ step/s. Since $J^+/J^-$ is not given in Ref.\cite{Walter_2012}, we used the value of $J^+/J^-$ obtained in Ref.\cite{nishiyama2002chemomechanical}, $J^+/J^-=221$. We set $F_d=3$pN \cite{M_ller_2010,vu2016discrete} and used the constraint $|d^+|+|d^-|=2.9$nm \cite{nishiyama2002chemomechanical}. By fitting to the run length distribution using our theory [Eq.(\ref{eq:run_dist})] along with the 4 constraints described above are used to  determine the model parameters at zero load, $k_0$, $k_0^+$, $k_0^-$, $\gamma_0$, and $K_T$. Subsequently, we used the data for average velocity vs load at different ATP concentrations in Ref.\cite{nishiyama2002chemomechanical} to obtain $d^+$ and $d^-$. The best fit parameters are listed in Table.1 and Table.2 in the main text for the 2HB waiting and 1HB waiting model, respectively.
\begin{figure}[]
\begin{center}
\includegraphics[width=\textwidth]{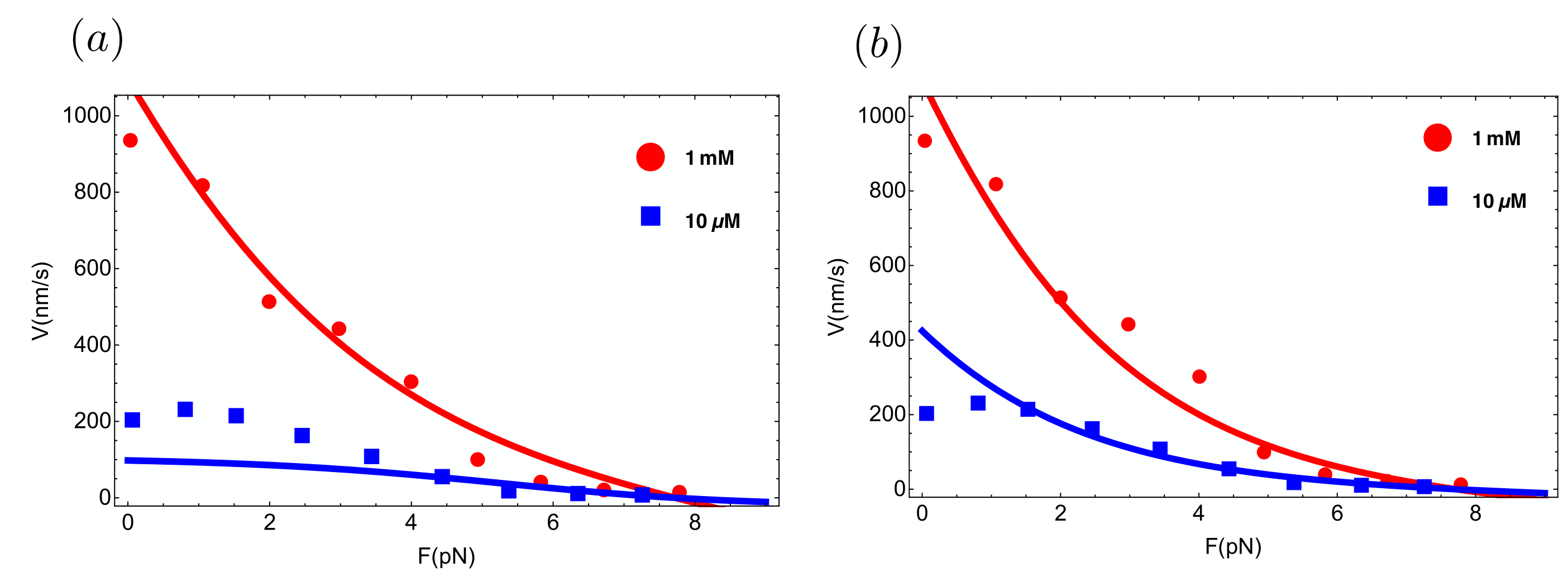}
\end{center}
\caption{\label{Fig:AveV}Fits of the  of average velocity as a function of load to the experiment \cite{nishiyama2002chemomechanical}. (a) 2HB waiting model [Fig.1(c) in the main text]. (b) 1HB waiting model [Fig.1(d) in the main text]. The agreement between theory and experiment is good.}
\end{figure}

\newpage{}
\bibliography{mybib}